\documentclass[twocolumn,preprint,]{aastex63}

\usepackage{amsmath,amstext}
\usepackage[T1]{fontenc}
\usepackage{apjfonts} 
\usepackage{graphicx}
\usepackage{natbib}
\usepackage{longtable}

\usepackage{microtype,longtable,verbatim,float,booktabs,graphicx}

\newcommand{\Ha}{\hbox{{\rm H}$\alpha$}}
\newcommand{\Lya}{\hbox{{\rm Ly}$\alpha$}}
\newcommand{\HavNII}{\hbox{({\rm H}$\alpha$+{\rm [N}\kern 0.1em{\sc II}{\rm ]})}}
\newcommand{\Hb}{\hbox{{\rm H}$\beta$}}

\newcommand{\SII}{\hbox{[\ion{S}{2}]}}
\newcommand{\NII}{\hbox{[\ion{N}{2}]}}
\newcommand{\OII}{\hbox{[\ion{O}{2}]}}
\newcommand{\OIII}{\hbox{[\ion{O}{3}]}}
\newcommand{\NeIII}{\hbox{[\ion{Ne}{3}]}}

\newcommand{\HII}{\hbox{{\rm H}\kern 0.1em{\sc II}}}

\newcommand{\OIIIHb}{\OIII/\Hb}
\newcommand{\NeOII}{\NeIII/\OII}

\newcommand{\Ms}{M_\star}

\begin{document}

\title{\large \bf CEERS Key Paper VIII: Emission Line Ratios from NIRSpec and NIRCam Wide-Field Slitless Spectroscopy at $z>2$}

\author[0000-0001-8534-7502]{Bren E. Backhaus}
\affil{Department of Physics, 196A Auditorium Road, Unit 3046, University of Connecticut, Storrs, CT 06269, USA}

\author[0000-0002-1410-0470]{Jonathan R. Trump}
\affil{Department of Physics, 196A Auditorium Road, Unit 3046, University of Connecticut, Storrs, CT 06269, USA}

\author[0000-0003-3382-5941]{Nor Pirzkal}
\affiliation{ESA/AURA Space Telescope Science Institute}

\author[0000-0001-6813-875X]{Guillermo Barro}
\affiliation{Department of Physics, University of the Pacific, Stockton, CA 90340 USA}

\author[0000-0001-8519-1130]{Steven L. Finkelstein}
\affiliation{Department of Astronomy, The University of Texas at Austin, Austin, TX, USA}

\author[0000-0002-7959-8783]{Pablo Arrabal Haro}
\affiliation{NSF's National Optical-Infrared Astronomy Research Laboratory, 950 N. Cherry Ave., Tucson, AZ 85719, USA}

\author[0000-0002-6386-7299]{Raymond C. Simons}
\affil{Department of Physics, 196A Auditorium Road, Unit 3046, University of Connecticut, Storrs, CT 06269, USA}

\author[0000-0002-4390-1816]{Jessica Wessner}
\affil{Department of Physics, 196A Auditorium Road, Unit 3046, University of Connecticut, Storrs, CT 06269, USA}

\author[0000-0001-7151-009X]{Nikko J. Cleri}
\affiliation{Department of Physics and Astronomy, Texas A\&M University, College Station, TX, 77843-4242 USA}
\affiliation{George P.\ and Cynthia Woods Mitchell Institute for Fundamental Physics and Astronomy, Texas A\&M University, College Station, TX, 77843-4242 USA}

\author[0000-0002-9921-9218]{Micaela B. Bagley}
\affiliation{Department of Astronomy, The University of Texas at Austin, Austin, TX, USA}

\author[0000-0002-3301-3321]{Michaela Hirschmann}
\affiliation{Institute of Physics, Laboratory of Galaxy Evolution, Ecole Polytechnique Fédérale de Lausanne (EPFL), Observatoire de Sauverny, 1290 Versoix, Switzerland}

\author[0000-0003-0892-5203]{David C. Nicholls}
\affiliation{Research School of Astronomy and Astrophysics, Australian National University, Canberra, ACT 2600, Australia}

\author[0000-0001-5414-5131]{Mark Dickinson}
\affiliation{NSF's National Optical-Infrared Astronomy Research Laboratory, 950 N. Cherry Ave., Tucson, AZ 85719, USA}

\author[0000-0001-9187-3605]{Jeyhan S. Kartaltepe}
\affiliation{Laboratory for Multiwavelength Astrophysics, School of Physics and Astronomy, Rochester Institute of Technology, 84 Lomb Memorial Drive, Rochester, NY 14623, USA}

\author[0000-0001-7503-8482]{Casey Papovich}
\affiliation{Department of Physics and Astronomy, Texas A\&M University, College Station, TX, 77843-4242 USA}
\affiliation{George P.\ and Cynthia Woods Mitchell Institute for Fundamental Physics and Astronomy, Texas A\&M University, College Station, TX, 77843-4242 USA}

\author[0000-0002-8360-3880]{Dale D. Kocevski}
\affiliation{Department of Physics and Astronomy, Colby College, Waterville, ME 04901, USA}

\author[0000-0002-6610-2048]{Anton M. Koekemoer}
\affiliation{Space Telescope Science Institute, 3700 San Martin Dr., Baltimore, MD 21218, USA}

\author[0000-0003-0492-4924]{Laura Bisigello}
\affiliation{Dipartimento di Fisica e Astronomia "G.Galilei", Universit\'a di Padova, Via Marzolo 8, I-35131 Padova, Italy}
\affiliation{INAF--Osservatorio Astronomico di Padova, Vicolo dell'Osservatorio 5, I-35122, Padova, Italy}

\author[0000-0002-6790-5125]{Anne E. Jaskot}
\affiliation{Department of Astronomy, Williams College, Williamstown, MA, 01267, USA}

\author[0000-0003-1581-7825]{Ray A. Lucas}
\affiliation{Space Telescope Science Institute, 3700 San Martin Drive, Baltimore, MD 21218, USA}

\author[0000-0003-1187-4240]{Intae Jung}
\affil{Space Telescope Science Institute, 3700 San Martin Drive Baltimore, MD 21218, United States}

\author[0000-0003-3903-6935]{Stephen M.~Wilkins} %
\affiliation{Astronomy Centre, University of Sussex, Falmer, Brighton BN1 9QH, UK}
\affiliation{Institute of Space Sciences and Astronomy, University of Malta, Msida MSD 2080, Malta}

\author[0000-0003-3466-035X]{{L. Y. Aaron} {Yung}}
\altaffiliation{NASA Postdoctoral Fellow}
\affiliation{Astrophysics Science Division, NASA Goddard Space Flight Center, 8800 Greenbelt Rd, Greenbelt, MD 20771, USA}

\author[0000-0001-7113-2738]{Henry C. Ferguson}
\affiliation{Space Telescope Science Institute, Baltimore, MD, USA}

\author[0000-0003-3820-2823]{Adriano Fontana}
\affiliation{INAF - Osservatorio Astronomico di Roma, via di Frascati 33, 00078 Monte Porzio Catone, Italy}

\author[0000-0002-5688-0663]{Andrea Grazian}
\affiliation{INAF--Osservatorio Astronomico di Padova, Vicolo dell'Osservatorio 5, I-35122, Padova, Italy}

\author[0000-0001-9440-8872]{Norman A. Grogin}
\affiliation{Space Telescope Science Institute, Baltimore, MD, USA}

\author[0000-0001-8152-3943]{Lisa J. Kewley}
\affiliation{Center for Astrophysics | Harvard \& Smithsonian, 60 Garden Street, Cambridge, MA 02138, USA}

\author[0000-0002-5537-8110]{Allison Kirkpatrick}
\affiliation{Department of Physics and Astronomy, University of Kansas, Lawrence, KS 66045, USA}

\author[0000-0003-3130-5643]{Jennifer M. Lotz}
\affiliation{Gemini Observatory/NSF's National Optical-Infrared Astronomy Research Laboratory, 950 N. Cherry Ave., Tucson, AZ 85719, USA}

\author[0000-0001-8940-6768]{Laura Pentericci}
\affiliation{INAF - Osservatorio Astronomico di Roma, via di Frascati 33, 00078 Monte Porzio Catone, Italy}

\author[0000-0003-4528-5639]{Pablo G. P\'erez-Gonz\'alez}
\affiliation{Centro de Astrobiolog\'{\i}a (CAB), CSIC-INTA, Ctra. de Ajalvir km 4, Torrej\'on de Ardoz, E-28850, Madrid, Spain}

\author[0000-0002-5269-6527]{Swara Ravindranath}
\affiliation{Space Telescope Science Institute, 3700 San Martin Drive, Baltimore, MD 21218, USA}

\author[0000-0002-6748-6821]{Rachel S. Somerville}
\affiliation{Center for Computational Astrophysics, Flatiron Institute, 162 5th Avenue, New York, NY, 10010, USA}

\author[0000-0001-8835-7722]{Guang Yang}
\affiliation{Kapteyn Astronomical Institute, University of Groningen, P.O. Box 800, 9700 AV Groningen, The Netherlands}

\author[0000-0002-4884-6756]{Benne W. Holwerda}
\affil{Physics \& Astronomy Department, University of Louisville, 40292 KY, Louisville, USA}

\author[0000-0002-8816-5146]{Peter Kurczynski}
\affiliation{Astrophysics Science Division, NASA Goddard Space Flight Center, 8800 Greenbelt Rd, Greenbelt, MD 20771, USA}

\author[0000-0001-6145-5090]{Nimish P. Hathi}
\affiliation{Space Telescope Science Institute, Baltimore, MD, USA}

\author[0000-0002-8018-3219]{Caitlin Rose}
\affil{Laboratory for Multiwavelength Astrophysics, School of Physics and Astronomy, Rochester Institute of Technology, 84 Lomb Memorial Drive, Rochester, NY 14623, USA}

\author[0000-0001-8047-8351]{Kelcey Davis}
\affiliation{Department of Physics, 196A Auditorium Road, Unit 3046, University of Connecticut, Storrs, CT 06269, USA}

\begin{abstract}

We use \textit{James Webb Space Telescope} Near-Infrared Camera Wide Field Slitless Spectroscopy (NIRCam WFSS)  and Near-Infrared spectrograph (NIRSpec)  in the Cosmic Evolution Early Release survey (CEERS) to measure rest-frame optical emission-line of 155 galaxies at $z>2$. The blind NIRCam grism observations include a sample of galaxies with bright emission lines that were not observed on the NIRSpec masks. We study the changes of the $\Ha$, $\OIIIHb$, and $\NeOII$ emission lines in terms of redshift by comparing to lower redshift SDSS and CLEAR samples. We find a significant ($>3\sigma$) correlation between $\OIIIHb$ with redshift, while $\NeOII$ has a marginal (2$\sigma$) correlation with redshift. We compare $\OIII/\Hb$ and $\NeIII/\OII$ to stellar mass and \Hb\ SFR. We find that both emission-line ratios have a correlation with \Hb\ SFR and an anti-correlation with stellar mass across the redshifts $0<z<9$. Comparison with MAPPINGS~V models indicates that these trends are consistent with lower metallicity and higher ionization in low-mass and high-SFR galaxies. We additionally compare to IllustrisTNG predictions and find that they effectively describe the highest $\OIIIHb$ ratios observed in our sample, without the need to invoke MAPPINGS models with significant shock ionizionation components.

\end{abstract}

  
\section{Introduction}\label{Intro}

Galaxy emission lines provide a wealth of information about galaxy formation and physical properties. Emission lines can be used to determine a galaxy's interstellar medium (ISM) conditions such as the metallicity, ionization, and density, as well as the physical properties such as star formation rate \citep{kenn12,Brin04} and dust attenuation \citep{card89,Calz1999,Reddy2016,Shap2023Balm}. One way to analyze emission lines is by comparing ratios of lines at similar wavelengths to gain information on galaxy ISM conditions. Picking emission-line ratios with similar wavelength makes the ratio less sensitive to dust attenuation. The most well-known emission-line ratio diagrams, BPT \citep{bald81} and VO87 \citep{Veil87},
compare $\OIII\lambda5007/\Hb$ with $\NII\lambda6583/\Ha$ or $\SII\lambda6583/\Ha$
to identify high ionization galaxies. These diagrams make use of the strongest emission lines in rest-frame optical spectra \citep{bald81,Veil87,Kauf03,Kewl06} to identify the dominant ionizing sources in galaxies.

Studying emission lines at different redshifts enables an understanding of how the physical conditions of galaxies change over cosmic time. At higher redshifts the strong rest-optical lines move to the near-IR. The cosmic star formation rate density is much higher at $z \sim 2$ than in the local Universe \cite{mada14}. Studies have found that galaxies at $z\sim 2$ have lower metallicities \citep{Stei14, henr13, Papo2022}, have higher ionization in the ISM, and exhibit $\alpha$-enhancement \citep{Stei16,shap19} than the local universe \citep{Liu08,shap15,strom18}, which is consistent with cosmic noon galaxies having higher star formation and more AGN. Due to the higher ionization, lower metallicity, and $\alpha$-element enhancement of star-forming galaxies at $z \sim 2$, the BPT and VO87 diagrams are not effective at distinguishing star-forming galaxies from AGN \citep{coil15,back22,Cleri2023a}. At redshift $z\sim2$ new emission-line diagrams were established such as the OHNO diagram, comparing $\NeIII\lambda3869/\OII\lambda3726+3729$ to $\OIIIHb$, to find high ionization sources instead \citep{back22,Zeim15}.

\begin{figure*}[t!]
\centering
\epsscale{0.9}
\plotone{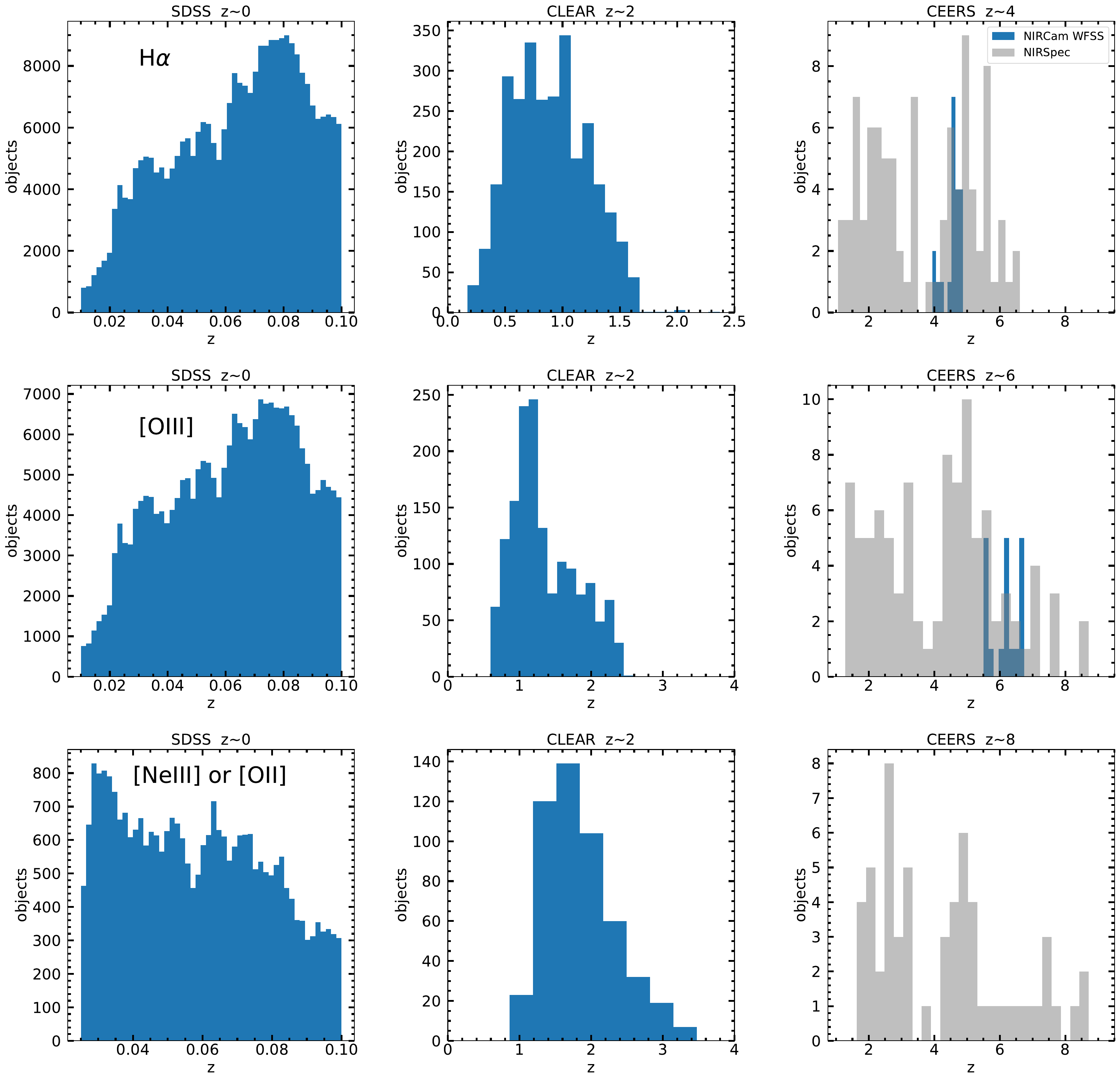}
\caption{Distribution of redshifts for each emission line sample: $Ha$(top), $\OIII$(middle), $\NeIII$ and $\OII$(bottom). These are named by there emission line which has a $SNR>3$, in the case of \NeIII\ and \OII\ either line may reach this requirement. The samples come from the SDSS(left), CLEAR (middle), and CEERS(right). In the right panel the gray histograms are the NIRSpec observations and blue are NIRCam WFSS.  
\label{fig:zsample}} 
\end{figure*} 

JWST now gives us access to the early universe, where the first galaxies and black holes are being formed \citep[e.g.,][]{Lars2023,Fink2023}. Which gives us the opportunity to study and understand the ISM conditions of galaxies at this early period of the universe. JWST observations will show how these galaxies assembled and evolved into the universe we see today. JWST near-IR spectroscopy allows us to view the rest-frame optical emission lines of galaxies at $2<z<9$.

Previous work has been done with NIRSpec spectroscopy from the JWST Early Release Observations (ERO) \citep{Ponto2022} of the lensing cluster SMACS J0723.3-7327 to investigate emission-lines of $z>5$ galaxies, however this uses a small sample of 3-6 galaxies. \cite{Schae22}, \cite{Curti23}, \cite{Trump22} studied the metallicities of these galaxies. \cite{Trump22}, \cite{Cleri2023}, and \cite{Brin22} went further and compared emission lines at $z>5$ to local samples, finding higher ($\NeOII$) in the early galaxies compared to the local ($z\sim0$) galaxies indicating higher ionization. Additionally, analysis has been done with the Cosmic Evolution Early Release Science (CEERS) survey program. The NIRSpec spectroscopy included in the CEERS survey has been used to gain a wealth of information such as: spectroscopically confirmed high-redshift galaxies \citep{Arra2023,Curtis2023,Fujimoto2023},  identifying and characterizing high-redshift AGN \citep{Koce2022, Lars2023}, study $\Lya$ emission \citep{Jung2023, Tang2023}, and studying ISM conditions as a function of galaxy properties such as stellar mass and star formation rate \citep{ Shap2023rel, Shap2023Balm}. \cite{Shap2023Balm} found there was not significant evolution between stellar mass and $\OIIIHb$ for galaxies above $z>3$, but their
stellar masses alone suggest subsolar metallicity. \cite{Shap2023Balm} also showed that the $z>5$ galaxies preference higher \Ha\ SFR when compared to \cite{Speagle2014} predictions. The ISM conditions of galaxies at $z>5$ were shown to have high ionization and low metallicity (>0.1 $Z_\odot$)\citep{Sand2023Ex}. \cite{Sand2023Te} makes use of $T_{e}$ based metallicity to calibrate strong-line metallicity estimators for $z>5$ galaxies \citep{Reddy2023}.

In this work, we use NIRCam WFSS and NIRSpec MSA spectroscopy taken as part of CEERS to investigate the rest-frame optical emission-line evolution and galaxy properties of $\sim$155 galaxies at $z>2$. In Section \ref{Data/Sample} we describe our data reductions and sample selection. Section \ref{Data/Sample} also establishes our comparison samples of galaxies at $z\sim0$ and $z \sim 2$. In Section \ref{SpecvCam} we compare galaxies in our NIRSpec and NIRCam WFSS samples. In Section \ref{Redshift} we use our three subsamples covering the epoch of reionization ($z>6$), cosmic noon ($z\sim2$), and the local universe ($z\sim0$) to study how each emission-line ratio evolves with redshift. Section \ref{Properties} describes the connections between galaxy properties and emission-line ratios at different redshifts. Section \ref{ISM} presents the ISM conditions inferred by the emission-line ratios. We summarize our results in Section \ref{Summary}. In this work, we assume a $\Lambda$ cold dark matter cosmology with $\Omega_{M}=0.3$, $\Omega_{\Lambda}=0.7$, and $H_{0}=70$ $km s^{-1}$ $Mpc^{-1}$ \citep{plan15}.

\section{Observational Data and Sample} \label{Data/Sample}

\subsection{JWST WFSS and MSA Spectroscopy}

Our parent galaxy sample comes from JWST observations taken by the CEERS program, ERS-1345 (PI: Steven Finkelstein). CEERS uses NIRCam WFSS and NIRSpec multi-object spectroscopy to cover $\sim$100 arcmin$^2$ of the Extended Groth Strip Hubble Space Telescope (HST) legacy field (EGS, \citealp{Davis2007}) which is covered by the CANDELS HST survey \citep{grog11, koek11}. Our paper focuses on the four pointings with NIRCam WFSS and six pointings with NIRSpec. CEERS has four NIRCam WFSS pointings that partially overlap with the CEERS NIRSpec observations allowing the emission line measurements of both instruments to be compared, allowing NIRSpec to be calibrated for slit losses.

\begin{figure*}[t!]
\centering
\epsscale{1.2}
\plotone{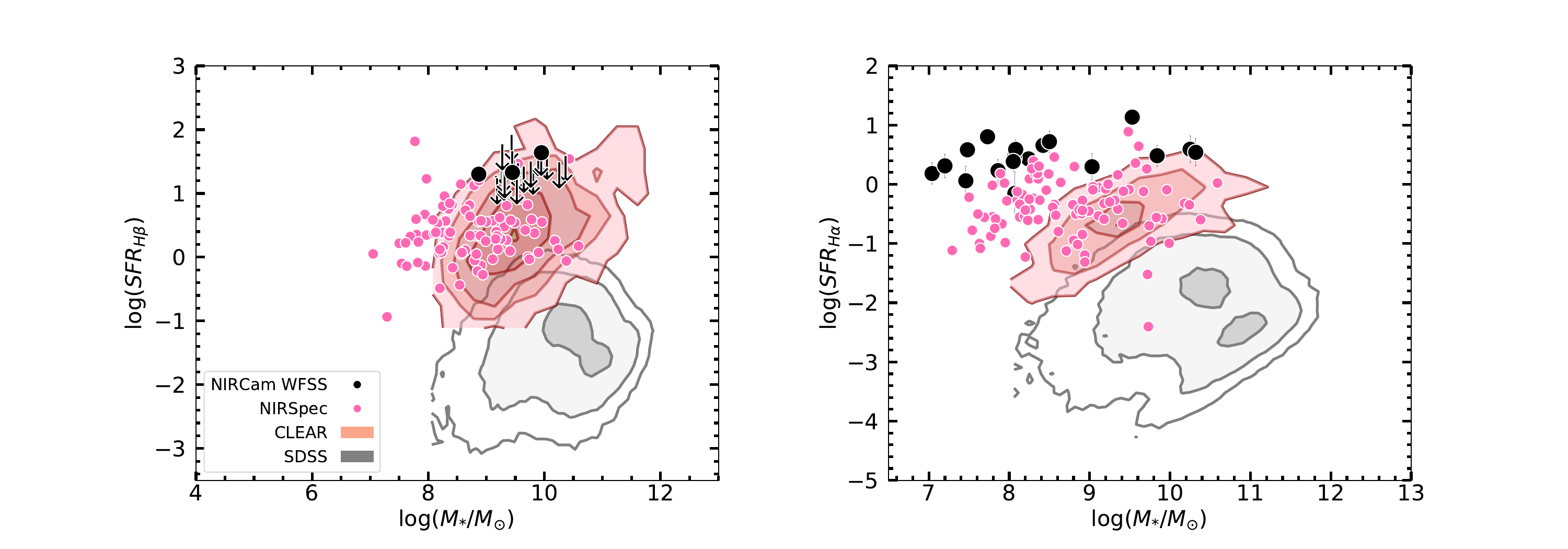}
\caption{Left: Distribution of \Hb\ SFR and stellar mass for our $\OIIIHb$ sample. Right: Distribution of \Ha\ SFR and stellar mass for our \Ha\ sample. In both the panels galaxies observed by NIRCam WFSS are represented by black points and arrows to represent lower limits,and galaxies observed by NIRSpec are represented by pink points. The red and gray contours represent the CLEAR and SDSS samples, respectively, with contour level indicating relative galaxy density of each sample.
\label{fig:samp_dist}} 
\end{figure*} 

\begin{figure*}[t!]
\centering
\epsscale{1.1}
\plotone{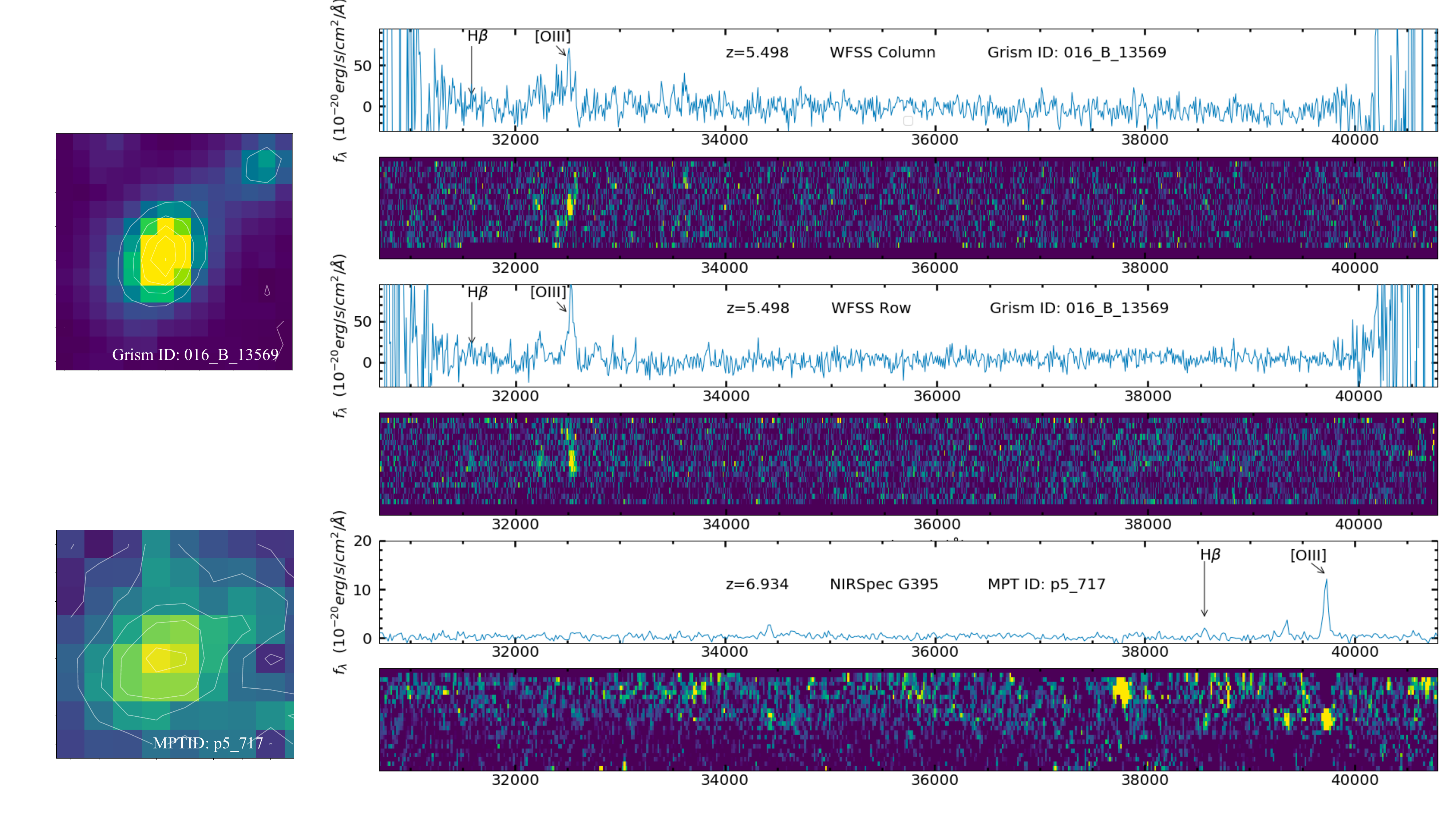}
\caption{Example of NIRCam WFSS direct image (left) with 1D/2D Column and Row WFSS spectra for an example galaxy, $016\_B\_13569$, that is in our $\OIIIHb$ sample. Vertical lines in the 1D spectra indicate the emission lines of interest. We also include the direct image and 1D/2D NIRSpec spectra of 717 from the G395M filter (bottom).  
\label{fig:exspec}} 
\end{figure*} 

\begin{deluxetable*}{|l|r|r|r|r|r|r|r|r|r|r|}[t]
\tablecaption{$\OIIIHb$ NIRCam Sample \label{tab:O3sample}}
\tablenum{1}
\tablecolumns{10}
\tablewidth{0pt}
\tablehead{\colhead{ID} & \colhead{R.A.} & \colhead{Dec.} & \colhead{z} &   \colhead{$\OIII$ Column} & \colhead{$\OIII$ Row} & \colhead{$\Hb$ Column} & \colhead{$\Hb$ Row} &  \colhead{$\log(\Ms)$} &  \colhead{log(SFR)} &  \colhead{$A_V$} 
\\
\colhead{} &  \colhead{deg} & \colhead{deg} & \colhead{} & \colhead{$10^{-18}erg/s/cm^2$} & \colhead{$10^{-18}erg/s/cm^2$} &
\colhead{$10^{-18}erg/s/cm^2$} &
\colhead{$10^{-18}erg/s/cm^2$} &
\colhead{[$M_\odot$]} &  \colhead{[$M_\odot$]/yr} &  \colhead{mag}}
\startdata
012\_B\_22258 & 214.929820 & 52.862205 & 5.635 &        49.37$\pm$ 6.33 &        41.37 $\pm$5.92 &      6.00$\pm$4.91 &      5.07 $\pm$4.91 &       10.05 &       2.16 &      0.8 \\
 013\_A\_8436 & 214.958411 & 52.843681 & 5.999 &        17.26 $\pm$2.02 &        17.06$\pm$2.03 &      3.65$\pm$ 2.07 &      2.33$\pm$1.87 &        9.37 &       0.45 &      0.0 \\
015\_B\_13732 & 214.987915 & 52.879438 & 6.177 &        44.09 $\pm$4.89 &        48.33$\pm$3.98 &      6.03$\pm$5.04 &      7.84$\pm$3.30 &        9.95 &       1.18 &      0.0 \\
015\_B\_19309 & 214.970312 & 52.881717 & 5.640 &         8.63$\pm$3.05 &         4.14$\pm$2.18 &      5.47$\pm$3.73 &      2.80$\pm$2.17 &        8.87 &       0.22 &      0.0 \\
012\_B\_30258 & 214.918284 & 52.879359 & 5.645 &        10.95$\pm$3.65 &        14.16 $\pm$3.47 &      2.85 $\pm$5.02 &      1.21$\pm$3.91 &        9.20 &       0.65 &      0.0 \\
013\_A\_15609 & 214.948684 & 52.856466& 6.530 &         3.47$\pm$1.29 &         7.49 $\pm$2.71 &      0.01 $\pm$1.19 &      1.52 $\pm$1.88 &        9.12 &       0.18 &      0.0 \\
015\_B\_19805 & 214.958370 & 52.875115& 6.170 &        34.25$\pm$12.26 &        33.86$\pm$5.80 &      2.43 $\pm$10.46 &      4.76$\pm$6.41 &        9.37 &       0.19 &      0.1 \\
016\_A\_13158 & 215.122563 & 52.973201& 6.380 &        42.02$\pm$5.94 &        49.80 $\pm$5.33 &      2.27 $\pm$4.11 &      4.32$\pm$ 5.29 &        9.89 &       1.31 &      0.0 \\
016\_A\_16127 & 215.127984 & 52.984951& 6.664 &        37.96$\pm$5.68 &        33.45$\pm$ 4.37 &      0.77 $\pm$4.58 &      1.70 $\pm$2.64 &        9.98 &       0.81 &      0.0 \\
016\_A\_18444 & 215.106518 & 52.975820& 6.174 &        12.27 $\pm$4.72 &        14.37$\pm$ 2.93 &      3.09$\pm$4.14 &      0.27$\pm$2.07 &        9.48 &       0.54 &      0.0 \\
012\_A\_27683 & 214.989007 & 52.919644& 6.670 &        11.81 $\pm$2.79 &        11.91$\pm$3.97 &      2.01$\pm$1.92 &      1.37$\pm$ 2.13 &        9.74 &       0.81 &      0.0 \\
013\_A\_10524 & 214.949129 & 52.843185& 6.716 &        18.91$\pm$ 4.62 &        16.83 $\pm$3.23 &     10.57 $\pm$4.10 &      0.52$\pm$3.69 &        9.83 &       1.82 &      0.8 \\
013\_B\_17930 & 214.869985 & 52.807034& 6.749 &        31.76 $\pm$6.47 &        32.30 $\pm$4.05 &      6.41$\pm$2.28 &      2.21$\pm$2.68 &       10.19 &       0.73 &      0.0 \\
 015\_B\_5666 & 215.000956 & 52.865869& 6.670 &        11.18$\pm$2.92 &         6.66$\pm$3.25 &      2.36$\pm$2.04 &      1.78$\pm$2.25 &        9.23 &       0.50 &      0.0 \\
012\_B\_28896 & 214.921870 & 52.876193& 5.630 &        14.94$\pm$6.96 &        12.38$\pm$3.83 &      2.35$\pm$3.91 &      4.41$\pm$ 1.50 &        9.24 &       0.14 &      0.0 \\
015\_A\_16826 & 215.032039 & 52.918960& 6.170 &        15.34 $\pm$ 3.71 &        14.03$\pm$3.33 &      4.51$\pm$4.90 &      2.12$\pm$1.99 &        9.70 &       0.76 &      0.1 \\
015\_A\_17952& 215.023039 & 52.915309 & 6.165 &         9.08$\pm$ 2.35 &         7.55 $\pm$3.45 &      0.84$\pm$3.02 &      2.67$\pm$1.51 &        9.25 &       0.31 &      0.0 \\
015\_B\_10107 & 214.987324 & 52.868911& 5.665 &        17.88$\pm$5.82 &        18.44$\pm$4.50 &      3.56$\pm$7.44 &      7.92$\pm$ 4.12 &        9.38 &       0.82 &      0.0 \\
016\_B\_13823 & 215.090344 & 52.951601& 5.501 &        18.82$\pm$8.36 &        29.20$\pm$4.27 &      5.69$\pm$13.39 &      4.50 $\pm$4.36 &        9.21 &       0.36 &      0.0 \\
\enddata
\caption{The emission line flux measurements for the \OIII\ NIRCam WFSS sample in units of $10^{-18} erg/s/cm^2$. Other columns show the spectroscopic redshift, the stellar mass, SED SFR and dust attenuation from the FAST SED fitting. The ID indicates the field$\_$panel$\_$WFSS ID number.}
\end{deluxetable*}

\begin{deluxetable*}{|l|r|r|r|r|r|r|r|r|}[t]
\tablecaption{$\Ha$ SFR NIRCam Sample \label{tab:Hasample}}
\tablenum{2}
\tablecolumns{9}
\tablewidth{0pt}
\tablehead{\colhead{ID} & \colhead{R.A.} & \colhead{Dec.} & \colhead{z} & \colhead{$\Ha$ Column} & \colhead{$\Ha$ Row} &  \colhead{$\log(\Ms)$} &  \colhead{log(SFR)} &  \colhead{$A_V$} 
\\
\colhead{} & \colhead{deg} &  \colhead{deg} &   \colhead{} & \colhead{$10^{-18}erg/s/cm^2$} & \colhead{$10^{-18}erg/s/cm^2$} &  \colhead{[$M_\odot$]} &  \colhead{[$M_\odot$]/yr} &  \colhead{mag}}
\startdata
012\_A\_25344 & 214.977646 & 52.903536 & 4.770 &       8.17 $\pm$1.79 &      21.61 $\pm$ 7.47 &        9.84 &       0.66 &      0.0 \\
012\_A\_25431 & 214.977566 & 52.903450& 4.550 &       9.36 $\pm$0.85 &       3.35 $\pm$1.88 &        7.46 &      -0.07 &      1.3 \\
012\_A\_31041 & 214.953995 & 52.907856& 4.550 &       8.52 $\pm$3.03 &       6.49$\pm$1.44 &        7.04 &      -0.53 &      0.8 \\
012\_A\_35810& 214.942294 & 52.919362 & 3.938 &      38.27 $\pm$8.28 &      50.02$\pm$6.51 &        7.73 &       0.20 &      1.3 \\
012\_A\_36545 & 214.937471 & 52.918283& 3.938 &      29.55$\pm$9.62 &      24.45 $\pm$10.11 &       10.25 &       0.24 &      0.1 \\
012\_B\_26391 & 214.927867 & 52.871022 & 4.805 &      12.42$\pm$1.67 &      10.74$\pm$3.26 &        8.24 &      -1.49 &      0.6 \\
012\_B\_29360& 214.917888 & 52.875555& 4.717 &      12.09$\pm$ 2.36 &      12.83$\pm$3.73 &        8.06 &      -1.95 &      0.3 \\
013\_A\_19410& 214.943410 & 52.864098 & 4.675 &      20.50 $\pm$5.27 &      12.32$\pm$5.67 &       10.32 &       1.38 &      0.8 \\
013\_A\_20628 & 214.941356 & 52.864855& 4.678 &       8.29$\pm$2.49 &       7.33 $\pm$2.36 &        7.86 &      -1.62 &      0.0 \\
013\_B\_13408 & 214.894682 & 52.812130& 4.876 &      14.12$\pm$3.88 &      18.39 $\pm$4.52 &        8.08 &      -2.73 &      0.3 \\
015\_A\_20196 & 215.022027 & 52.920785& 4.540 &      14.11$\pm$3.47 &       6.73 $\pm$2.95 &        9.03 &      -9.90 &      2.9 \\
015\_B\_16325 & 214.978092 & 52.879514& 4.545 &      16.27$\pm$4.19 &      21.61 $\pm$3.85 &        7.48 &      -0.09 &      1.2 \\
015\_B\_17048& 214.985862 & 52.886907 & 4.546 &       9.39$\pm$3.17 &      10.73$\pm$ 3.36 &        7.20 &      -0.33 &      1.3 \\
 016\_A\_5887 & 215.151987 & 52.974048& 4.480 &      29.21$\pm$ 4.90 &      18.21 $\pm$ 2.09 &        8.42 &       0.61 &      1.9 \\
 016\_A\_8530 & 215.149544 & 52.978974& 4.525 &       1.93$\pm$1.59 &       6.29 $\pm$ 0.15 &        8.07 &      -8.80 &      0.7 \\
016\_B\_15982& 215.059038 & 52.936442  & 4.280 &      65.93$\pm$16.13 &      89.81$\pm$2.93 &        9.53 &       0.97 &      0.0 \\
016\_B\_18194& 215.079996 & 52.956800& 4.745 &      30.26$\pm$7.73 &      17.87$\pm$5.60 &        8.50 &      -0.46 &      0.5 \\
 016\_B\_8414 & 215.083373 & 52.931987& 4.117 &      11.18 $\pm$ 3.61 &      20.14 $\pm$ 3.73 &        8.05 &      -1.68 &      0.7 \\
\enddata
\caption{The \Ha\ flux measurements from NIRCam WFSS, spectroscopic redshift, the stellar mass, SED SFR and dust attenuation from the FAST SED fitting. The ID indicates the field$\_$panel$\_$WFSS ID number.}
\end{deluxetable*}

\begin{deluxetable*}{|l|r|r|r|r|r|r|r|r|r|r|}[t]
\tablecaption{NIRSpec Sample \label{tab:NIRSpec}}
\tablenum{3}
\tablecolumns{10}
\tablewidth{0pt}
\tablehead{ \colhead{R.A.}& \colhead{Dec.}& \colhead{z} &   \colhead{$\OIII$} &    \colhead{$\Hb$} & \colhead{$\Ha$} &  \colhead{$\NeIII$} & \colhead{$\OII$}&  \colhead{$\log(\Ms)$} &  \colhead{log(SFR)} &  \colhead{$A_V$}\\
 \colhead{deg} &\colhead{deg} &\colhead{} &\colhead{ $10^{-18}ergs/s/cm^2$} &\colhead{ $10^{-18}ergs/s/cm^2$} &\colhead{ $10^{-18}ergs/s/cm^2$} &\colhead{ $10^{-18}ergs/s/cm^2$} &\colhead{ $10^{-18}ergs/s/cm^2$} &\colhead{[$M_\odot$]} &\colhead{$[M_\odot]$/yr} & \colhead{mag}} 
\startdata
214.957160 & 52.872372 & 3.228 &       2.57 $\pm$0.31 &    0.54 $\pm$0.43 &     1.57$\pm$0.21 &        1.82 $\pm$ 0.52 &      1.22$\pm$0.47 &        8.91 &       0.18 &      0.0 \\
214.959997 & 52.831169  & 4.900 &       3.09 $\pm$0.19 &    0.25$\pm$0.12 &     1.20 $\pm$0.08 &      -99.00 $\pm$0.00&    -99.00 $\pm$0.00 &        9.11 &       0.67 &      0.0 \\
214.893181 & 52.882484 &3.000 &       1.83$\pm$0.22 &    0.78$\pm$ 0.35 &     1.49 $\pm$ 0.27 &        2.05$\pm$0.54 &      2.10$\pm$ 0.59 &        8.80 &       0.06 &      0.0 \\
214.943900 & 52.850052 &5.001 &      65.25$\pm$19.57 &    8.19$\pm$0.88 &    30.26$\pm$ 0.84 &        3.60 $\pm$ 0.45 &     11.17$\pm$0.84 &        9.48 &       0.74 &      0.0 \\
214.941496 & 52.850565 &2.540 &      10.00 $\pm$4.94 &    5.29 $\pm$0.52 &    18.24$\pm$ 1.76 &        0.56$\pm$ 1.13 &     10.69$\pm$0.80 &        9.03 &       0.48 &      0.0 \\
 214.907360 & 52.844535 &2.010 &      15.23 $\pm$1.67 &    1.83$\pm$1.01 &     5.80$\pm$0.25 &        1.05$\pm$ 0.77 &      2.53$\pm$1.02 &        7.54 &      -0.86 &      0.0 \\
214.898480 & 52.861709 & 1.922 &       1.35$\pm$0.64 &    0.79 $\pm$ 0.58 &     4.22$\pm$0.53 &        0.52$\pm$ 229.54 &      4.63$\pm$0.60 &        9.76 &       0.84 &      0.3 \\
214.909604 & 52.880284 &2.144 &       4.79 $\pm$0.37 &    1.19 $\pm$0.42 &     2.85$\pm$ 0.18 &        0.36$\pm$ 0.46 &      2.59$\pm$0.43 &        8.85 &       0.41 &      0.0 \\
214.966546 & 52.846672 & 2.136 &      19.14 $\pm$3.83 &    3.99$\pm$0.54 &    14.20$\pm$ 0.76 &        1.07 $\pm$.92 &     11.29$\pm$1.42 &        9.18 &       0.25 &      0.0 \\
214.940182 & 52.836026 &1.699 &       9.83$\pm$0.97 &    6.28$\pm$1.29 &    25.11$\pm$1.09 &        2.14 $\pm$1.29 &     16.69$\pm$2.59 &       10.18 &       1.19 &      0.8 \\
\enddata
\caption{The emission-line fluxes from the NIRSpec sample. The galaxy IDs are formatted as pointing\_ID and redshift. Emission-line fluxes are reported in units of $10^{-18} ergs/s/cm^2$. The SFR and dust attenuation measurements are produced from SED fitting. A machine-readable version of the full table is available.}
\end{deluxetable*}

The four NIRCam WFSS pointings use the F356W filter to cover $3.14-3.98$~\micron\, including a suite of rest-frame optical lines for $0<z<9$ galaxies with a total exposure time of 2490 sec between the two orthogonal gratings in the WFSS. The WFSS spectra have spectroscopic resolving power R$\sim$1,600 at $\sim 4 \micron$. The details of the NIRCam wide-field slitless spectrograph (WFSS) are described by \cite{Green2017}. 

The six NIRSpec spectroscopy pointings use the G140M/F100LP, G235M/F170LP and G395M/F290LP filters, spanning 1-5~$\micron$. Four of these pointings were observed in medium resolution, R$\sim$1000, and in prism R$\sim$100. Two pointings had light-leak failures in the prism observations and we use only the medium-resolution grating observations (the failed prism observations were rescheduled and observed later but are not used in this work). The details of the NIRSpec instrument are described by \citet{jakobsen22}. Further information on the programs used in the CEERS extractions and fits can be read in \cite{Bagley2023} and \cite{Arra2023}.

\subsubsection{Data Reduction and Sample Selection}

Information on the reductions of the NIRSpec data can be found in Arrabal Haro et al., (in prep.). The reduction's main steps are as follows. The NIRSpec data is processed with the STScI Calibration Pipeline version 1.8.5 \footnote{\url{https://jwst-pipeline.readthedocs.io/en/latest/index.html}} and the Calibration Reference Data System (CRDS). The \texttt{calwebb\_detector1} pipeline module was used to reduce the uncalibrated images by applying the correction for ``snowballs'' events caused by cosmic rays, the 1/f noise correction, and doing a saturation check. An improved correction of the ``snowball'' events\footnote{\url{https://jwst-docs.stsci.edu/data-artifacts-and-features/snowballs-and-shower-artifacts}} are applied to the \texttt{jump} step. This step also creates the count-rate maps (CRMs).

The generated CRMs were then passed through the \texttt{calwebb\_spec2} pipeline to create two-dimensional (2D) cutouts of the slitlets. The \texttt{calwebb\_spec2} pipeline performs the background subtraction by using a 3-nod pattern, corrects the flat-fields, implements the wavelength and photometric calibrations and resamples the 2D spectra to correct the distortion of the spectral trace. 

The \texttt{calwebb\_spec3} pipeline stage creates the final 2D spectra, by combining the images of the three nods. Our data uses the default slit loss correction settings to create this spectra. The 1D spectra are created by extracting from the 2D spectra using customized apertures. These custom apertures are visually defined for targets to maximize the signal-to-noise ratios (SNRs).

The NIRSpec flux uncertainties are underestimated by a factor of $\sim2$, determined by comparing he normalized median absolute deviation (NMAD) of the flux to the  median of the flux uncertainty for each source. We correct for this by increasing the flux uncertainty of each spectrum by the ratio of the NMAD of the flux to the median flux uncertainty, $\mathrm{NMAD}(f) / \mathrm{median}(\sigma_f)$.

The NIRCAM WFSS data were first processed with the Stage 1 STScI pipeline to apply bias and dark corrections and perform on-the-ramp fitting to detect and remove cosmic ray impacts. A broad filter F356W flat-field was applied to the data as it was shown to reduce pixel-to-pixel variation. Finally, proper world coordinate information was added using the AssignWCS() Pipeline task.
Spectra were extracted from these processed datasets following the Simulation Based Extraction \cite{pirz17}. We used an extraction catalog derived from the CEERS F356W mosaic. 
Spectral contamination was modeled and subtracted using pixel-level SEDs for each source computed from the CEERS F277W, F356W, and F444W mosaics.
Finally, the dispersed background was subtracted as a combination of pre-launch synthetic dispersed F356W background and of an additional row (column) model of the residuals in the GRISMR (GRISMC) observations. 
This results in the modeling and extractions of 24,000 spectra. Continuum is detected in sources down to approximately $\sim25.75$ AB magnitude.

The NIRCam WFSS flux uncertainties are underestimated by a factor of $\sim3$ in the row dispersion and $\sim4$ in the column dispersion. This was determined and corrected using the same method with the ratio of NMAD of the flux to the median of the uncertainty used for the NIRSpec sample.

Our photometric redshifts used for sample selection are obtained using the same method outlined in \cite{Fink2022b}. Briefly, the multi-band SEDs were fit using the \textsc{EAZYpy} \citep{bram08} software package.  Probability distribution functions (PDFs) are created by fitting non-negative linear combinations of templates to the observed data. The templates used are a set of 12 FSPS “$tweak\_fsps\_QSF\_12\_v3$” \cite{conr10}, and six additional templates to cover bluer colors. These bluer color templates are shown by \cite{Lars2023} to improve the photometric redshift fits for z$>9$ galaxies. These templates use stellar population models created with BPASS \cite{2009Eldridge} with 5$\%$ solar metallicities and young stellar populations, $\log$(age/yr)=6,6.5,7.

\subsubsection{Sample Selection}
To extract the emission line flux from both NIRCam WFSS, we first use photometric redshifts to constrain our sample to find galaxies of interest. Our constraints were determined by which redshifts would have our desired emission lines land in the F356W filter. This would give redshift ranges of $3.7<z<5.1$ for \Ha, a $5.4<z<7$ range for \OIII, and $7.4<z<9.3$ for $\OII$. We increase the range of each redshift bin to account for galaxies that may have an inaccurate photometric redshift, using ranges
of $3.2<z<5.5$ for \Ha, a $4.9<z<7.5$ range for \OIII, and $6.9<z<9.8$ for $\OII$ to be further inspected. The photometric redshift overlap between \OIII\ and \OII\ is due to adding $\Delta z = 0.5$ to select galaxies. Meanwhile for NIRSpec we use the photometric redshift constraint of $1.6<z<6.8$ as this covers all emission lines spanning the G140M, G235M, and G395M filters. We determined the spectroscopic redshift for NIRCam and NIRSpec for each source using the best-fit line center of the brightest emission line in each spectrum, usually \OIII$\lambda$5008 or $\Ha$ as they are the brightest lines in the spectra. To measure emission line fluxes we find the best-fit Gaussian function (and associated uncertainties) using a Levenberg-Marquardt least-squares method implemented by the scipy \texttt{curvefit} python code. 

Our NIRSpec sample has a total of 118 galaxies, 93 galaxies in the \Ha\ sample, 96 galaxies in the \OIII\ sample, and 59 in the \OII\ sample. These galaxies follow the same sample selection process as NIRCam WFSS, requiring a $S/N>3$ for at least one emission line. The last sample includes \NeIII$\lambda$3869 and the blended \OII$\lambda$3726+3729 doublet, which is accessible up to z$\sim$8. Figure \ref{fig:exspec} shows examples of NIRSpec
1D and 2D spectra on the bottom right panels.

We do not dust correct the emission lines of our samples because both \Ha\ and \Hb\ are not available for all galaxies in the sample. The emission-line ratios used in this work are not affected by dust attenuation due to the fact that the emission-line pairs are close in wavelength and are nearly equally affected by dust. We note that 62$\%$ of our \OIIIHb\ galaxies between NIRSpec MSA and NIRCam WFSS at $z>4$ have an SED $A_V<0.2$, so we expect the \Hb\ SFR estimates to be minimally effected by dust. Meanwhile only 48$\%$ of our \Ha\ galaxies at $z>4$ have an SED $A_V<0.2$, and only three galaxies in the NIRCam WFSS \Ha\ sample have SED $A_V<0.2$. Galaxies affected by dust will have a lower limit for \Hb\ or \Ha\ SFR.

SFR is calculated from either the \Hb\ or \Ha\ emission line, depending on redshift, by following the \citet{kenn12} SFR relation for \Ha\ and $\Ha/\Hb=2.86$ (assuming Case B recombination, $T=10^4$~K, and $n_e=10^4$~cm$^{-3}$; \citealp{oste89}):

\begin{equation} \label{Eq:SFRHb}
  \log({\rm SFR})[M_\odot/\mathrm{yr}] = \log[L(\Hb)] - 40.82\
\end{equation}

\begin{equation} \label{Eq:SFRHa}
  \log({\rm SFR})[M_\odot/\mathrm{yr}] = \log[L(\Ha)] - 41.27
\end{equation}

Due to the \Hb\ and \Ha\ lines not being dust corrected these SFR are a lower limit.

We create two WFSS subsamples of galaxies based on a signal-to-noise ratio of $\mathrm{S/N}>3$ detection for at least one line of interest when the two orients are coadded. The first includes galaxies with $\mathrm{S/N}>3$ in $\Ha$ at z$\sim$4. The second includes galaxies with $\mathrm{S/N}>3$ \OIII$\lambda$5007 emission lines at z$\sim$6. We note there are no galaxies with \NeIII$\lambda$3868 or \OII$\lambda$3728 detected at S/N$>$3 in the CEERS NIRCam WFSS observations due to the low observation depth. We visually inspect the filter image, 1D, and 2D spectra of galaxies selected for these samples to ensure the emission line is detected in both orients. An example of one galaxy inspected and included in the samples is shown in Figure \ref{fig:exspec}. 

The sample selection and inspection results in
18 galaxies with \Ha\ in the redshift range $4<z<5$ and 19 galaxies with \OIII\ in the redshift range $5.5<z<7$ from the NIRCam WFSS.

Tables \ref{tab:O3sample} and \ref{tab:Hasample} present the source IDs, spectroscopic redshifts, emission line flux, stellar mass, SED SFR, and dust attenuation measurements of the NIRCam WFSS \OIII\ and \Ha\ galaxy samples. Our NIRSpec sample is shown in Table \ref{tab:NIRSpec} showing the ID, redshift, emission line fluxes. Figures \ref{fig:zsample} and \ref{fig:samp_dist} show the redshift distribution, stellar mass and SFR of both samples.

\begin{figure*}[t!]
\centering
\epsscale{1.2}
\plotone{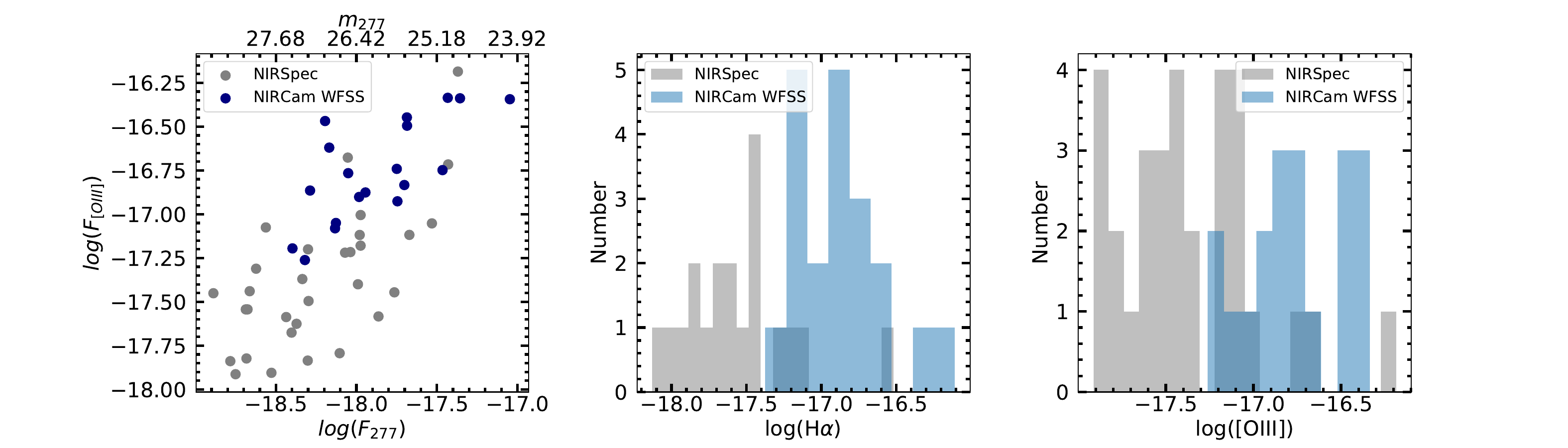}
\caption{Left: Comparison of the NIRSpec and NIRCam WFSS measurements of F277W continuum flux and \OIII\ flux. Comparison between emission line flux from NIRSpec (gray histogram) and NIRCam WFSS (blue histograms) for Middle: \Ha\ emission and Right: \OIII\ emission. Both these emission lines were seen in both instruments. A blind survey like NIRCam WFSS includes galaxies that have been missed in the targeted NIRSpec observations: the NIRCam WFSS galaxies tend to have stronger emission lines.
\label{fig:CamvSpec}} 
\end{figure*} 

\subsubsection{Stellar Mass}

Stellar masses and dust attenuations in our NIRCam WFSS and NIRSpec sample come from fitting the optical and NIR SEDs from the long-wavelength NIRCam filters using \textit{FAST} \cite{krie09}.  These models assume stellar population synthesis models used by \cite{bruz03}, following the initial mass function (IMF) defined by \cite{chab03}, a \cite{calz00} dust attenuation law, and a delayed exponential star formation history.

\subsection{Lower Redshift Comparison Samples}

We established two $z\sim2$ comparison samples to cover the peak of cosmic star formation \citep{mada14} and supermassive black hole growth \citep{aird10}. A SDSS $z\sim0$ comparison sample was also established to cover the local universe \citep{Brin04,Kauf03a,Kauf03b}.

\subsubsection{SDSS $z \sim 0$ Sample}

The $z\sim0$ comparison sample is created from the Sloan Digital Sky Survey \citep[SDSS;][]{york00} Data Release 10 \citep{Ahn14}. The SDSS data set used a 2.5 m telescope at Apache Point Observatory to cover 14,555 deg$^2$ in the sky with $R \sim 2000$ over $3800<\lambda<9200$ \AA\ \citep{smee13}.

Emission-line measurements and redshifts for the SDSS data set are computed by \cite{bolt12}, using a stellar template to correct the continuum for stellar absorption. Stellar masses are estimated by \cite{Mont16} from the broadband $ugriz$ SDSS photometry using a grid of templates made from the FSPS stellar population synthesis code \citep{conr09}. These templates assume a \cite{krou01} IMF and fit for the dust attenuation following \cite{Char00} and \cite{calz00}.

The low-redshift, z$\sim$0, comparison sample was selected using the same $\mathrm{S/N}>3$ line detection thresholds as for the CEERS and CLEAR samples. These selection criteria results in 284,523 galaxies with $\Ha$, 231,999 galaxies  with $\OIIIHb$, and 27,847 galaxies with $\NeOII$. We note this SDSS sample has no cuts on whether a galaxy is an AGN or SF.

\subsubsection{CLEAR z$\sim$1.5 Sample}

Our $z\sim2$ comparison sample comes from the CANDELS Ly$\alpha$ Emission at Reionization (CLEAR) survey \citep{simons2023} and \textit{HST} near-IR spectroscopy with the G102 and G141 grisms taken as part of the 3D-HST program \citep{momc16,bram12,vand11}. 

We select the CLEAR comparison sample using the same $\mathrm{S/N}>3$ line detection thresholds as for the CEERS sample, visually inspecting the direct image, 1D and 2D spectra to remove galaxies with contaminated spectra. This gives us 2890 galaxies with $\Ha$, 1534 galaxies with $\OIII$, and 505 in our $\OII$ sample. Due to this sample's low spectral resolution ($R\sim100$), \Ha\ is blended with \NII$\lambda$6583+6548. This blending causes the \Ha\ fluxes of CLEAR galaxies to effectively be lower limits, although most $z \sim 2$ galaxies have $\NII/\Ha \ll 1$ (e.g., \citealp{shap15}).

\subsubsection{MOSDEF z$\sim3$ Sample}

We also compare to the stacked line-ratio measurements from the MOSFIRE Deep Evolution Field (MOSDEF) observations used in \cite{sanders21} at redshift $z\sim3.3$. This sample was observed by the Multi-Object Spectrometer For Infrared Exploration (MOSFIRE; \cite{McLean2012}) on the 10~m Keck~I telescope, observing 48.5 nights over a four year period to obtain rest-frame optical spectra of $1.4<z<3.8$ galaxies \citep{krie15}. This survey covers the AEGIS, COSMOS, and GOODS-N fields. In order to cover multiple emission lines, two to three filters are used in the survey to observe H-band selected galaxies. MOSDEF adopts a slit width of 0.7 arcseconds, which results in a spectral resolution of R = 3400, 3000, 3650 and 3600 for Y, J, H, and K, respectively. 

We use the stacked line-ratio measurements from \citet{sanders21}, which were calculated from $\sim$750 galaxies at z $\sim$ 2.3 and $\sim$375 galaxies at  z$\sim$ 3.3. For a full description of the MOSDEF survey design and data reduction, see \cite{krie15}.

\begin{figure*}[t!]
\centering
\epsscale{1.1}
\plotone{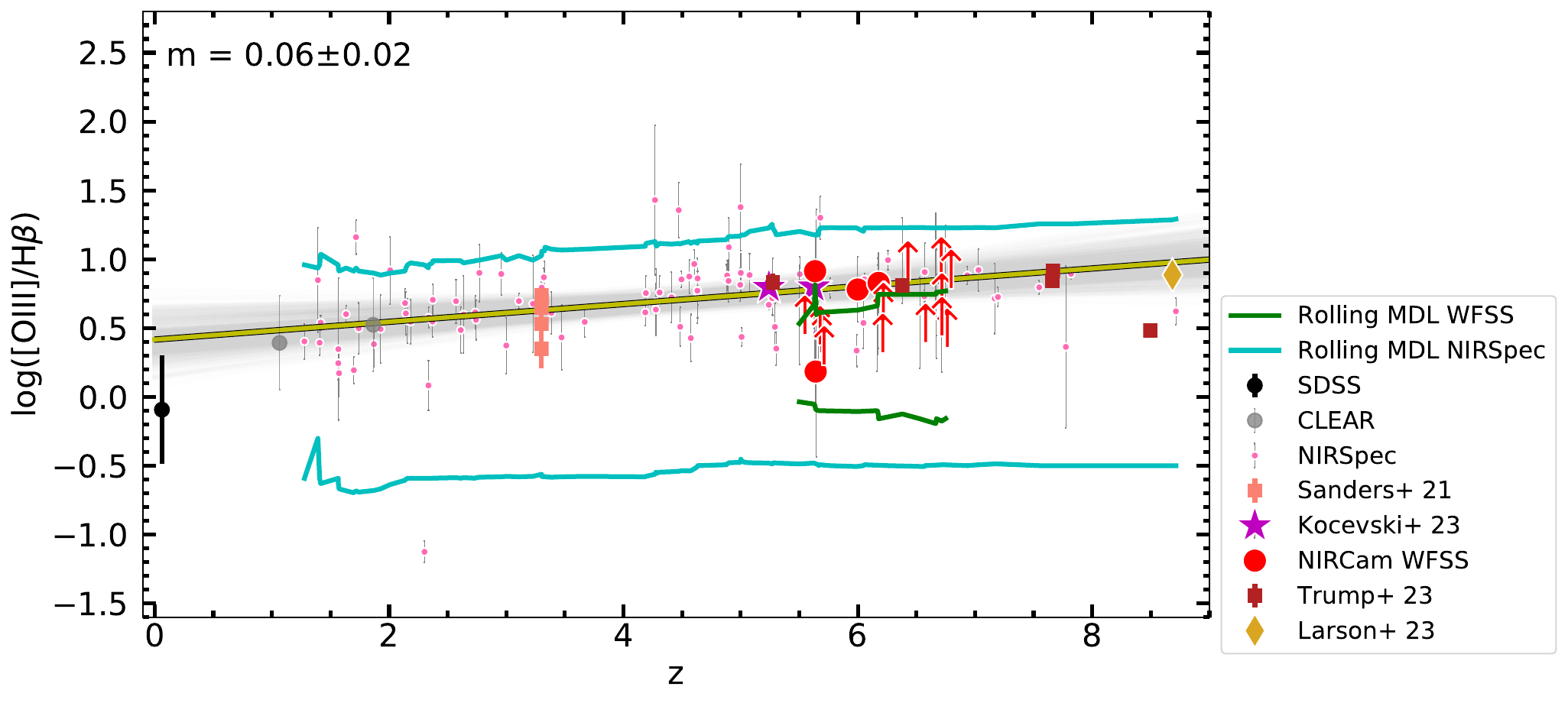}
\caption{The $\log{\OIIIHb}$ emission-line ratio versus redshift. The black point is the median redshift and $\log{\OIIIHb}$ of the SDSS sample \cite{york00}, with the error bars representing the standard deviation of the sample. The gray points and pink squares are from the CLEAR and MOSDEF samples respectively \citep{simons2023, sanders21}. The SMACS observations are marked as crimson squares. NIRSpec AGN from CEERS are purple stars. The NIRSpec and WFSS samples are represented by pink points and red points and arrows respectively. The yellow line is the linear fit to the NIRSpec, and NIRCam WFSS data. The cyan lines are the rolling median detection limit (MDL) to the best-fit line for the NIRSpec galaxies which indicates the range of galaxies we could see based on our S/N cutoffs. This increase of $\OIIIHb$ with redshift is also shown in the linear fit line, with a slope of (0.06$\pm$0.02).  
\label{fig:O3Hb_zcom}} 
\end{figure*} 

\section{NIRCam WFSS and NIRSpec MSA Comparison} \label{SpecvCam}

The NIRCam WFSS is a blind survey that allows us to gain access to a sample of galaxies that may be missed by targeted NIRSpec observations, which require pre-selection of galaxies based on redshift and brightness. In the right panel of Figure \ref{fig:CamvSpec} the NIRSpec $z>5$ galaxies are represented by gray circles, while the blue circles represent the combined column and row measurements of NIRCam WFSS. The NIRCam WFSS galaxies occupy a similar region of continuum measurements, however these galaxies tend to have higher \OIII\ emission. 

Due to this difference in sample selection, the NIRCam WFSS provides an opportunity to view different types of galaxies. These galaxies also have differences in the emission line measurements. Figure \ref{fig:CamvSpec} also shows a comparison of the flux measurements of \OIII\ and \Ha\ between NIRCam WFSS and NIRSpec. These emission lines were chosen as they have S/N$>3$. NIRSpec is represented by the gray histogram and the combined measurements from both the Row and Column dispersions for NIRCam WFSS are shown in the blue histograms. This comparison shows that NIRCam WFSS tends to include galaxies with brighter emission line fluxes.
This indicates that the CEERS NIRSpec observations are missing galaxies with more extreme emission lines. This should be kept in mind in our following sections when comparing NIRSpec and NIRCam WFSS galaxies.

\section{Redshift Evolution of Emission Line Galaxies} \label{Redshift}

We measure the galaxy emission-line ratios with redshift by comparing CEERS galaxies with $z\sim2$ galaxies from CLEAR and MOSDEF and with $z\sim0$ galaxies from SDSS.

\begin{figure*}[tbp]
\centering
\epsscale{1.1}
\plotone{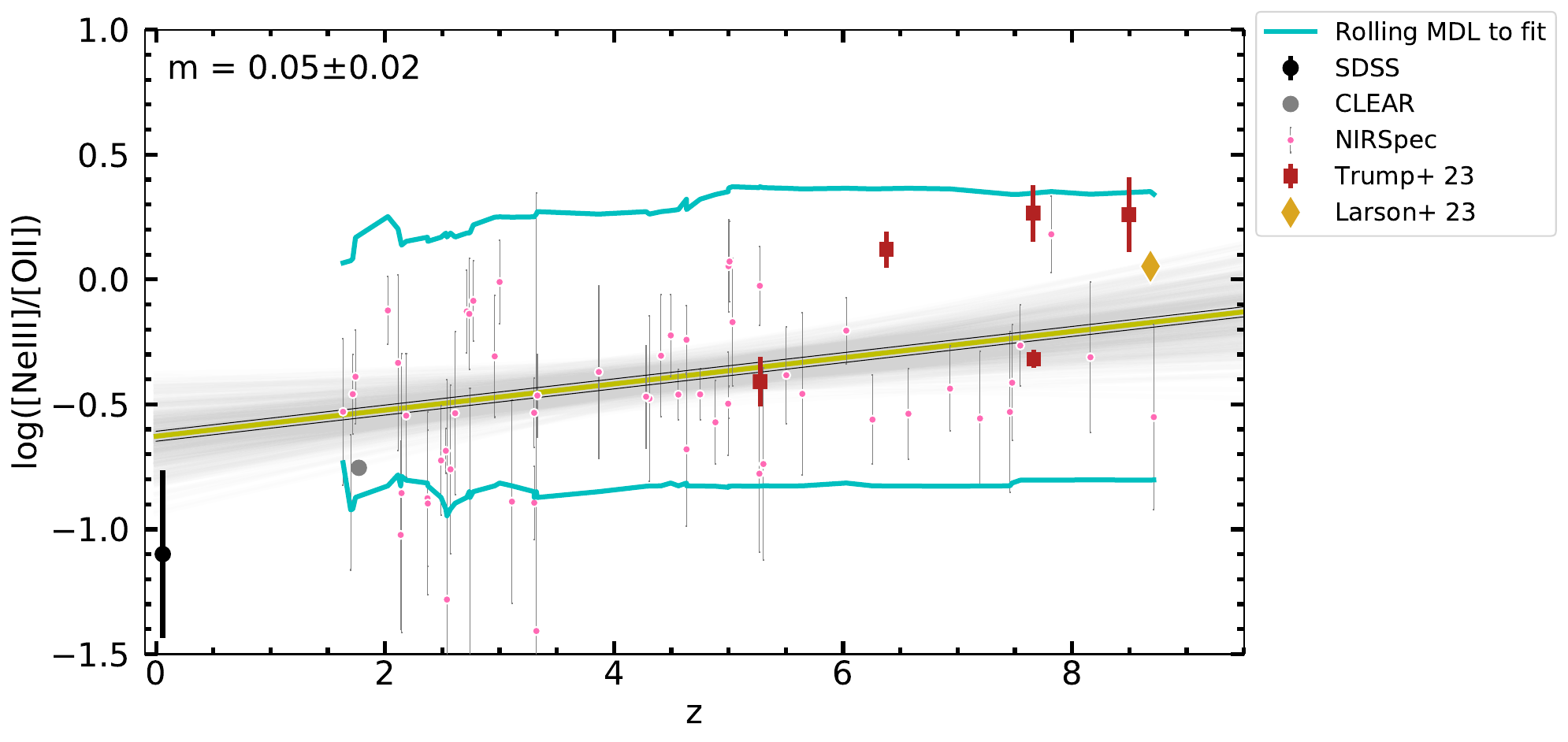}
\caption{The $\log{\NeOII}$ emission-line ratio versus redshift. The black point is the median redshift and $\log{\NeOII}$ of the SDSS sample \cite{york00}. The gray point is a median value from CLEAR \cite{simons2023}. The pink points and red squares are from the CEERS NIRSpec and SMACS samples respectively. The NIRSpec $\NeOII$ line ratio has a 2.5$\sigma$ slope with redshift, (0.05$\pm$0.02), as shown by the yellow best fit line. However, if you remove the SDSS and CLEAR samples we find the rest of the sample has a slope of 0.05 with a 2$\sigma$ certainty.
\label{fig:NeO2_z}} 
\end{figure*}

In Figure \ref{fig:O3Hb_zcom} we plot the \OIIIHb\ emission-line ratios against redshift. The black point is the median value from the SDSS sample, the gray points median values from two redshift bins from the CLEAR sample, and the \cite{sanders21} redshift 3.3 binned data are shown as pink squares. The pink circles are our NIRSpec sample and the red points are the CEERS sample. Arrows in our CEERS WFSS sample represent lower limits in the emission-line ratio due to undetected $\Hb$. Other high redshift galaxies from SMACS ERO NIRSpec observations \citep{Trump22} and broad-line AGN from \cite{Koce2022} and \cite{Lars2023} are marked as crimson squares, purple stars, and gold diamonds respectively. There is a 0.49 dex increase in $\OIIIHb$ between the SDSS sample and CLEAR, see \cite{back22} which carefully matched the two samples to avoid differences in sample selection for details and discussion. We also find that the $z>5$ CEERS sample has 0.33 dex higher $\OIIIHb$ than the CLEAR sample, but we note this may be due to the difference in the luminosity selection. A \texttt{linmix} linear fit to the line ratios measured from CEERS NIRSpec and NIRCam WFSS observations indicates a shallow but significant (3$\sigma$) increase of \OIIIHb\ with redshift, this fit includes the lower limits. Many of the NIRCam WFSS line ratios are lower limits due to undetected \Hb\ lines: these limits are generally consistent with the measured NIRSpec line ratios but is also consistent with a steeper increase of \OIIIHb\ with redshift.

The median detection limits (MDL) are created by taking the upper and lower limit of each galaxy. The resulting plotted lines are the difference between the detection limits and the fit. This line represents the lowest signal that can be observed with a 1$\sigma$ detection of the line flux. We note that our NIRSpec MSA data at high redshifts are well separated from the rolling MDL, indicating there is no selection bias for our high redshift sample. Thus the small increase of \OIIIHb\ ratio with redshift is not likely to be a simple result of a changing detection limit with redshift.

Figure \ref{fig:NeO2_z} shows the \NeOII\ ratio with redshift, using the same notation as Figure \ref{fig:O3Hb_zcom}. The $\NeOII$ ratio also increases with redshift. There is a 0.34 dex increase between the SDSS and CLEAR samples. This is different from the results from \cite{back22} which found $\NeOII$ had a 0.2 dex difference between SDSS and CLEAR. This may be because \cite{back22} required a S/N>1 in both \OII\ and \NeIII\ where we only require S/N>3 for $\OII$ for both SDSS, CLEAR, and CEERS samples.  
There is also a 0.37 dex increase between CLEAR and galaxies with z>5. This is smaller than the 0.5 dex increase between CLEAR and SMACS reported by \cite{Trump22}, however \cite{Trump22} only had 5 galaxies above $z>5$. There is only a marginal correlation between the $\NeOII$ ratio and redshift in the CEERS samples with a slope of $0.05\pm0.02$. The rolling MDL indicates the CEERS $z>2$ sample may be affected by the detection limit caused by the S/N$>1$ detection threshold for \NeIII, such that there might exist $z>1.5$ galaxies with lower $\NeOII$ that are undetected. On the other hand, the measured line ratios are well-separated from the upper detection limit and there appears to be a genuine lack of high-$\NeOII$ galaxies at $z>2$.

\begin{figure}[h!]
\centering
\epsscale{1.3}
\plotone{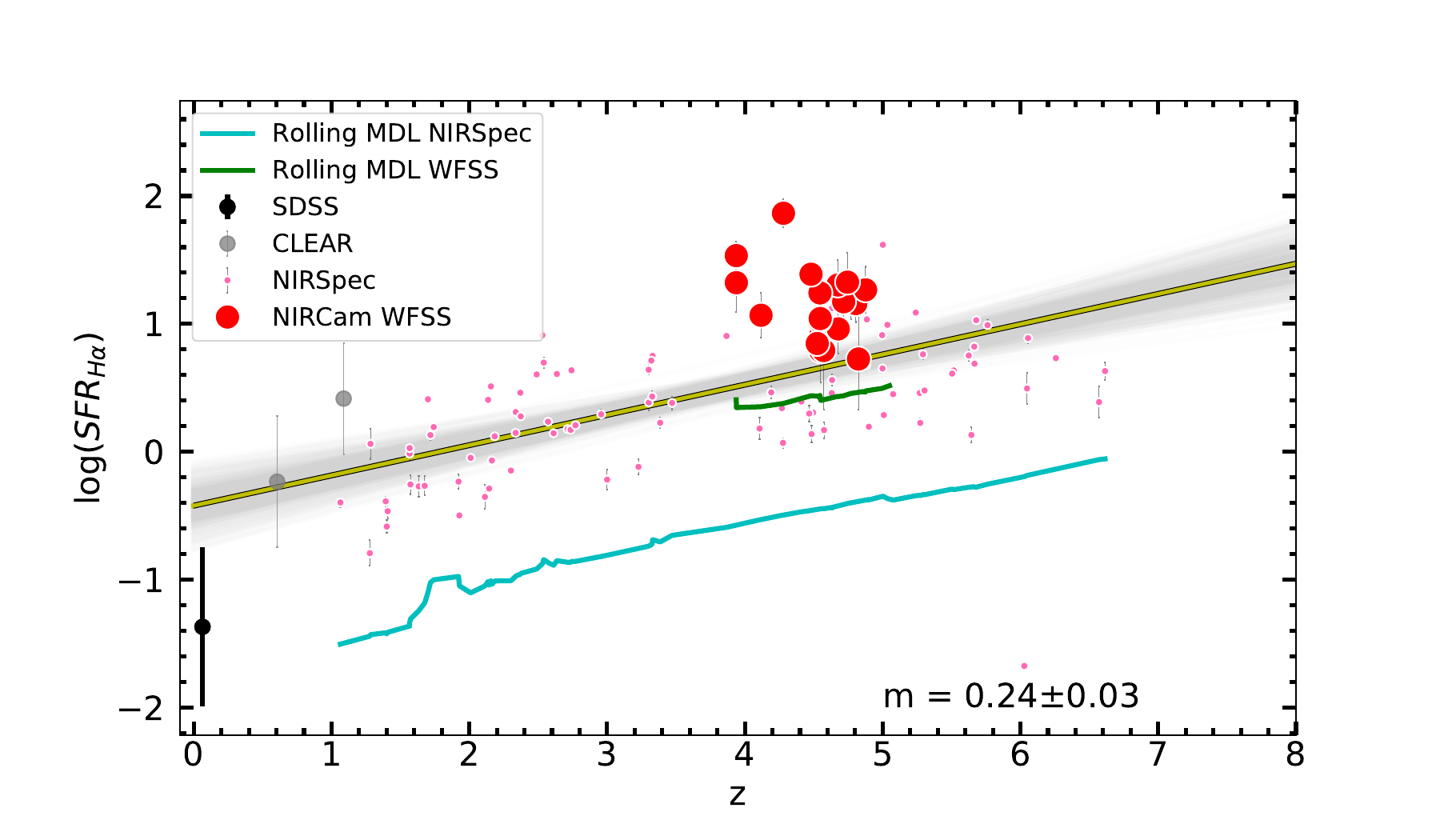}
\caption{The $\Ha$ SFR versus redshift. The black point is the median value of the SDSS sample \cite{york00}. The gray point is a median value from CLEAR \cite{simons2023}. The pink points and red circles are from the CEERS NIRSpec and NIRCam WFSS samples respectively. The cyan and green lines represent the lower detection of the NIRSpec and NIRCam WFSS data respectively. There is a 1.5 dex increase in the median $\Ha$ SFR measurements from SDSS to CLEAR and a
0.5 dex increase in the median $\Ha$ SFR measurements between CLEAR and $z>5$ galaxies. The yellow best fit line to the NIRSpec and NIRCam $\Ha$ SFR a 8$\sigma$ slope with redshift, (0.24$\pm$0.03). 
\label{fig:Ha_z}} 
\end{figure} 

Finally, we show the relationship between $\Ha$ SFR with redshift in Figure \ref{fig:Ha_z}. This sample is not dust corrected as we do not have a pair of hydrogen lines in all observations and the SED-based $A_V$ may be unreliable for estimating the nebular attenuation. We note that the \Ha\ NIRCam WFSS galaxies have higher $A_V$ measured by their SED fitting than the other samples. This would push the galaxies to higher $\Ha$ SFR. We again see a correlation of $\Ha$ SFR with redshift, With a 1.5~dex increase between SDSS and CLEAR, and a 0.5~dex increase between CLEAR and CEERS galaxies with $z>5$. We see a 0.7 dex difference between the NIRCam WFSS and NIRSpec galaxies between $3.8<z<5$ due to the different detection limits of each instrument.

The relationships between emission lines and galaxy properties and ISM conditions are further explored
in Section \ref{Properties} and \ref{ISM} of the paper.

\section{Emission Line Properties with Galaxy Stellar Mass and SFR}\label{Properties}

We will now investigate how the emission-line ratios correlate to galaxy propterties such as stellar mass and SFR.

\begin{figure*}[tbp]
\centering
\epsscale{1.1}
\plotone{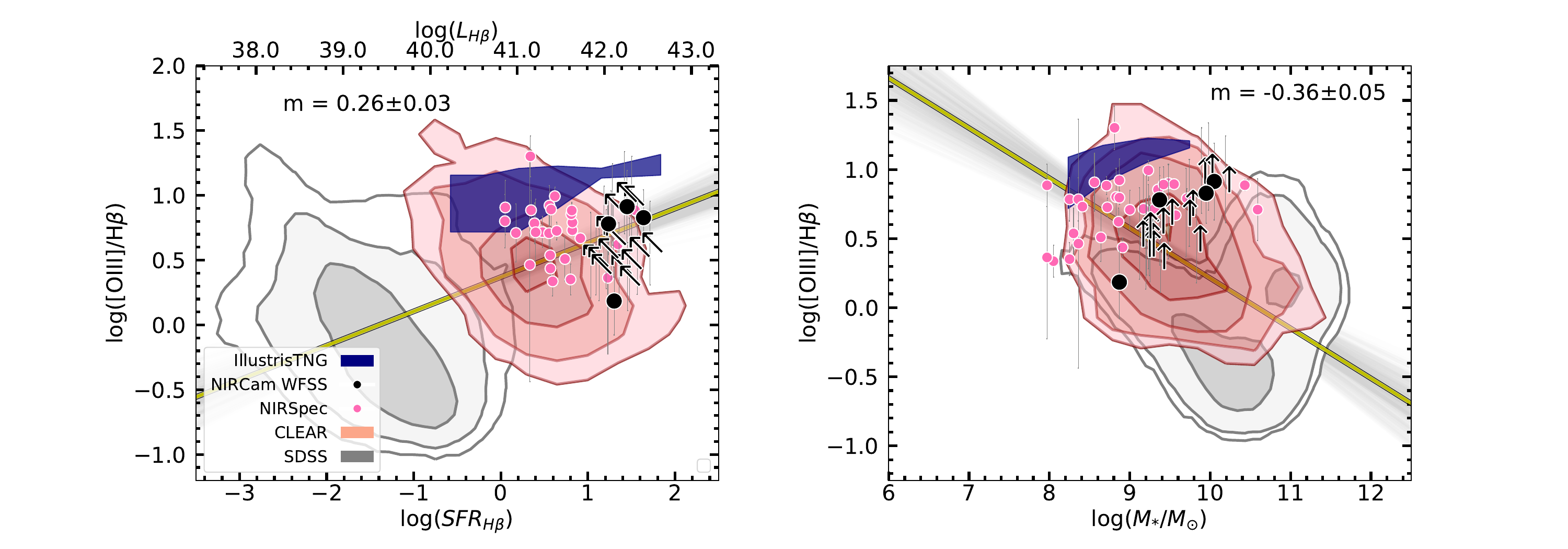}
\caption{Left: The relationships between the $\OIIIHb$ emission-line ratio and the galaxy \Hb\ luminosity and \Hb\ SFR. Right: The relationships between the $\OIIIHb$ emission-line ratio and stellar mass. In both panels the gray and red contours are the SDSS and CLEAR sample respectively. The black points and arrows are $\OIIIHb$ measurements and limits from NIRCam WFSS observations. The yellow fit line is to the CEERS $z>5$ sample and randomly selected SDSS and CLEAR galaxies matching the size of the CEERS sample. All the IllustrisTNG simulations are shown as the purple region. 
\label{fig:O3_prop}} 
\end{figure*} 

\begin{figure*}[btp]
\centering
\epsscale{1.1}
\plotone{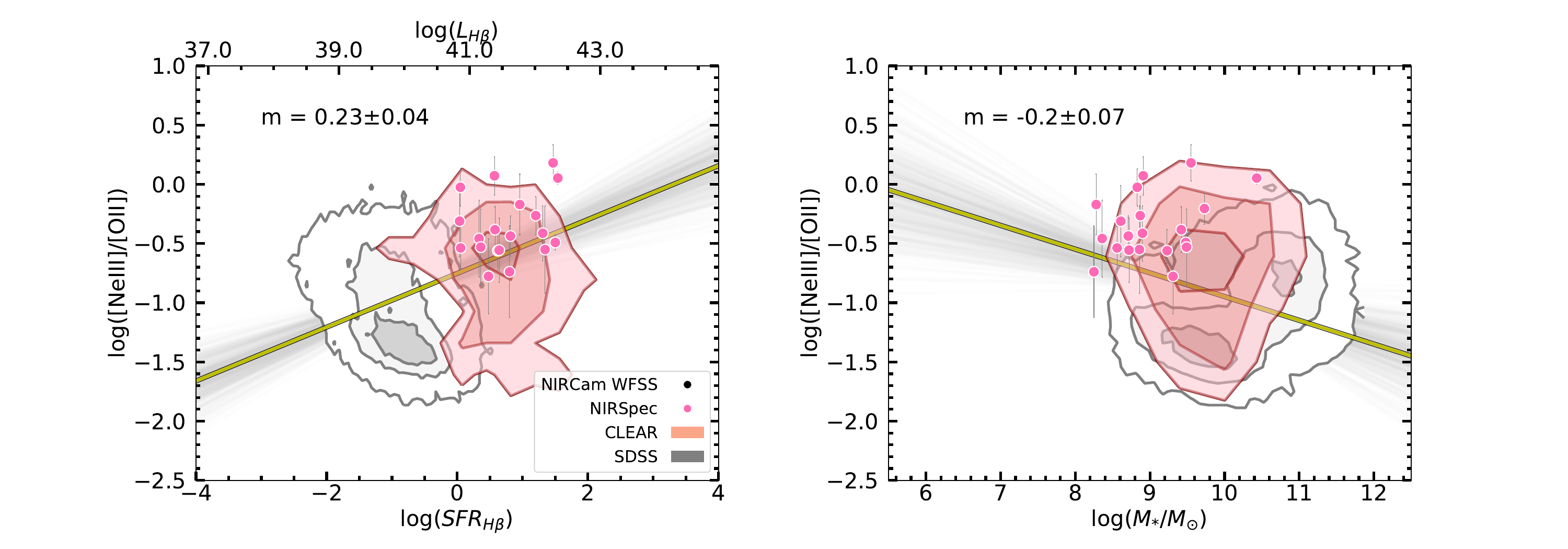}
\caption{Left: The relationships between the $\NeOII$ emission-line ratio and the galaxy \Hb\ luminosity and \Hb\ SFR. Right: The relationships between the $\NeOII$ emission-line ratio and stellar mass. In both panels  the gray and red contours are the SDSS and CLEAR sample respectively. The black points and arrows are $\OIIIHb$ measurements and limits from NIRCam WFSS observations. The yellow fit line is to the CEERS $z>5$ sample and randomly selected SDSS and CLEAR galaxies matching the size of the CEERS sample. There is a significant ($>3\sigma$) relationship between $\NeOII$ and \Hb\ SFR and a marginal relationship ($2.8\sigma$) with stellar mass.
\label{fig:Ne3_prop}} 
\end{figure*}

\begin{figure*}[tbp]
\centering
\epsscale{1}
\plotone{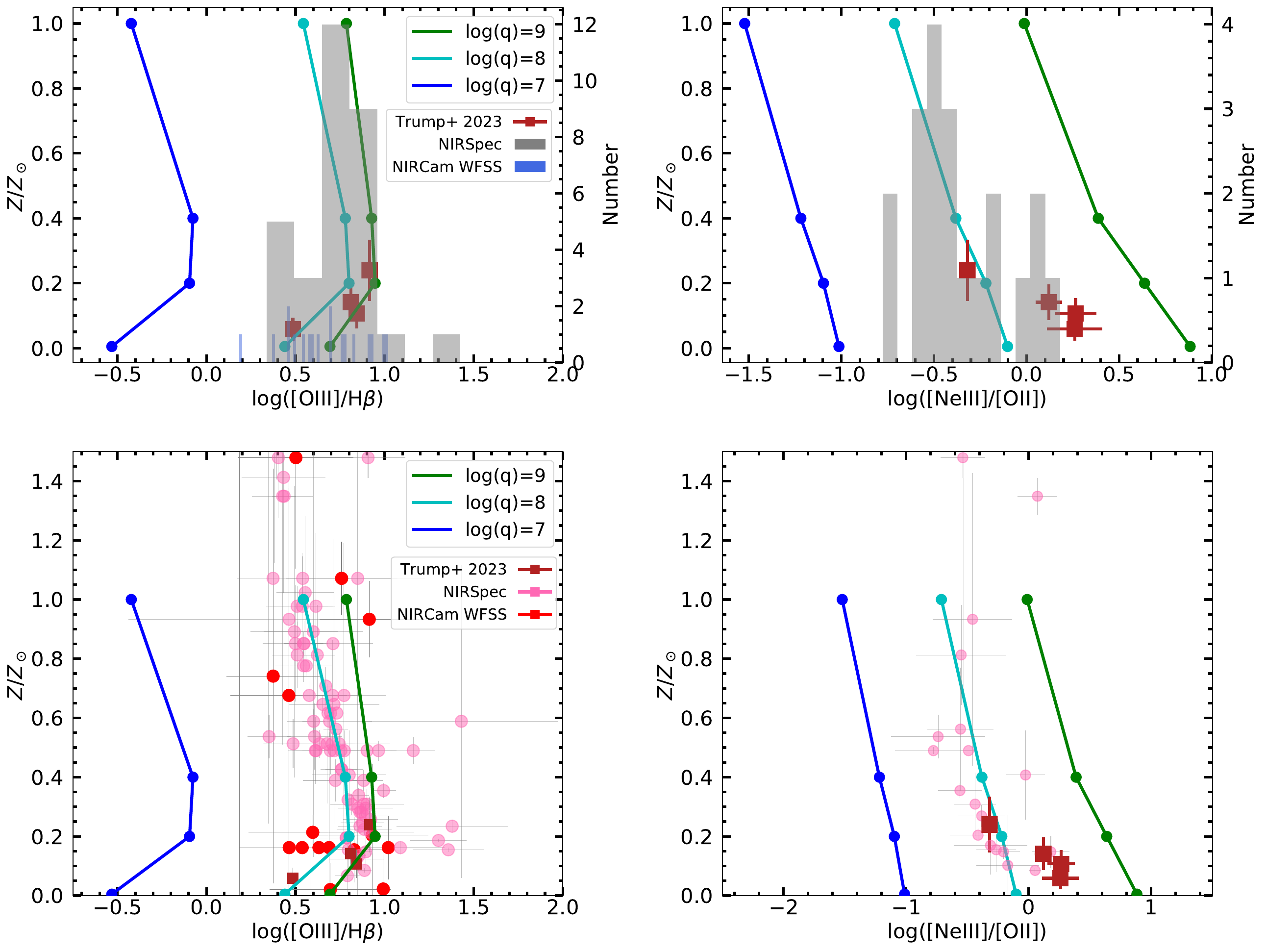}
\caption{Top Left: Comparison of $\OIIIHb$ measurements of our sample to MAPPINGS~V models of ISM ionization and metallicities. The gray histogram represents the distribution of the NIRSpec measurements of $\OIIIHb$ while the blue histogram represents the NIRCam WFSS sample, both CEERS samples do not have a measured metallicity. We also include the four SMACS galaxies which have metallicity measurements using the coronal \OIII$\lambda$4364. Top Right: Comparison of $\NeOII$ measurements of our sample to MAPPINGS V models of the ISM's ionization and metallicity. The gray histogram represents the distribution of the NIRSpec measurements of $\NeOII$. Bottom: Distribution of the CEERS $z>5$ sample where the metallicities are derived using the strong-line calibrations provided in \cite{Sand2023Te}.
\label{fig:Mapping}} 
\end{figure*} 

Figure \ref{fig:O3_prop} relates the $\OIIIHb$ ratio to $\Hb$ SFR and stellar mass. Figure \ref{fig:Ne3_prop} similarly compares $\NeOII$ to $\Hb$ SFR and stellar mass. Equation \ref{Eq:SFRHb} was used to derive $\Hb$ SFR, this is a proxy for the unobscured galaxy SFR. Here our $z\sim0$ SDSS and $z\sim2$ CLEAR samples are represented by the gray and red contours, respectively. Our NIRCam WFSS galaxies are represented by black points for well measured galaxies and black arrow for lower limits. NIRSpec galaxies are shown as pink points. In Figure \ref{fig:O3_prop} we include the median value of the Illustris similations as the purple line. We calculate a best-fit line from the python \texttt{linmix} linear regression package, using subsamples of SDSS galaxies and CLEAR galaxies the same size of the CEERS sample.

We additionally compare our observed \OIIIHb\ to the simulated ratios of \citet{Hirsch2023}. These model line ratios are built on the IllustrisTNG simulations (reference), applying both Cloudy \citep{Ferl13} and MAPPINGS~V \citep{Suth18} photoionization models to the simulated galaxies. These simulated galaxies include a mix of nebular emission contributions from star-forming \HII\ regions, post-AGB stars, shocks, and AGN narrow-line regions. The details of these model line ratios are described by \citet{Hirsch2017} and \citet{Hirsch2022}. We compare to model line ratios from simulated $z \sim 6$ galaxies with $\log(M_\star/M_\sun) \sim 9$ and $\log(\mathrm{SFR}/[M_\sun/\mathrm{yr}]) \sim 0.5$, matching the median stellar mass and SFR of the observed galaxies. The range of line ratios from these simulated galaxies are shown as purple shading in Figure \ref{fig:O3_prop}.

The $\OIIIHb$ emission-line ratio has a significant ($>$3$\sigma$) correlation with \Hb\ SFR with a slope of $0.26\pm0.03$, and a significant anti-correlation with stellar mass with a slope of $-0.33\pm0.05$. Our $z>5$ NIRCam WFSS and NIRSpec sample are on average 0.5~dex higher in \Hb\ SFR than the CLEAR galaxies. The IllustrisTNG simulated line ratios are effective at reproducing the highest \OIIIHb\ ratios observed in our sample, but most of our observed galaxies have lower \OIIIHb\ ratios than the simulated galaxies of similar galaxy mass and SFR. We attempted to do multiple linear regression on our sample to detangle the relationship between the $\OIIIHb$ emission line with redshift, \Hb\ SFR, and stellar mass however due to the sample size and the differences between the NIRSpec and NIRCam WFSS samples produced unreliable results. A larger sample size would be needed for further analysis.

\begin{figure*}[tbp]
\centering
\epsscale{0.9}
\plotone{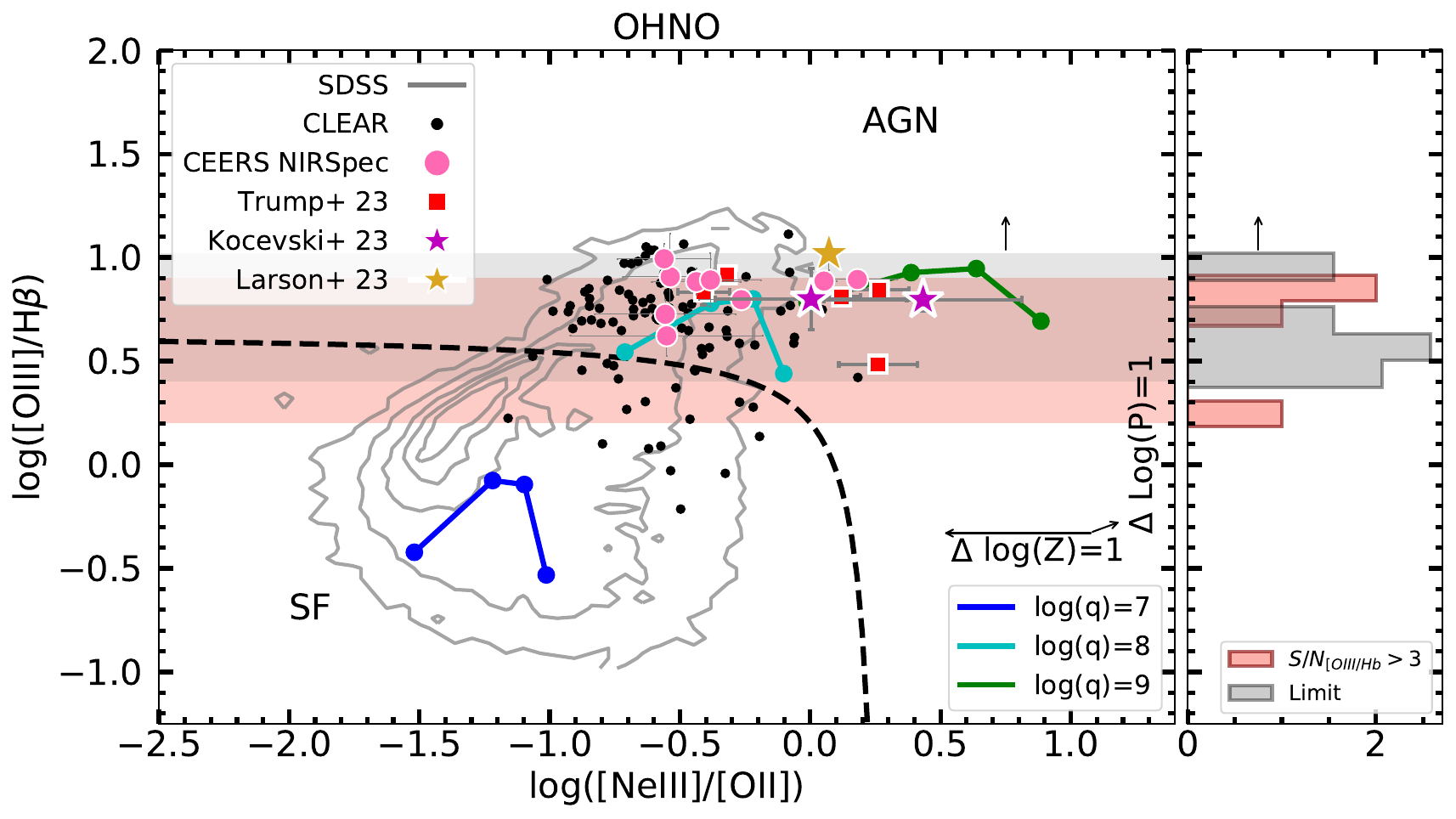}
\caption{Theoretical OHNO line ratio predictions from the MAPPINGS~V models (colored lines) compared to the low-redshift SDSS galaxies (gray contours), cosmic noon CLEAR galaxies (black points), SMACS galaxies (red stars), CEERS AGN (purple stars) and CEERS NIRSpec galaxies (pink points). Due to the constraint of the filter only $\OIIIHb$ can be observed for the NIRCam WFSS observations. These galaxies are represented by the histogram in the right hand panel, the gray histogram represents the NIRCam WFSS measurements with a lower limit and red histogram are galaxies with a S/N measurement$>3$ for both emission lines. The arrow on this histogram indicates that the true \OIIIHb\ values may be higher. The red histogram in the panel represents the well measured galaxies from the NIRCam WFSS, this sample is also represented by the pink bar across the panel. Model ionization is indicated by the color of the line, and metallicity decreases from left to right for each set of connected model points. Inset vectors indicate the direction and amplitude of 1~dex increases in metallicity and pressure. The black dashed line is the empirical AGN/SF dividing line defined for $z\sim2$ galaxies in \cite{back22}. The CEERS data set is best described by moderate/high ionization and a moderate metallicity, while CLEAR is described by lower ionization and more moderate metallicity.
\label{fig:O3Hb_ISM}} 
\end{figure*}

The $\NeOII$ ratio has a significant correlation with \Hb\ SFR with a slope of $0.23\pm0.04$, which is similar to $\OIIIHb$. Similar to $\OIIIHb$, $\NeOII$ has a marginal (2$\sigma$) anticorrelation to stellar mass with a slope of $-0.2\pm0.07$, but is about three times less significant what was found between stellar mass and $\OIIIHb$.

The anti-correlation between $\log(\OIII/\Hb)$ and stellar mass has been shown in \citet{dick16} and \citet{Kash19}, and is due to lower metallicity and higher ionization in galaxies with higher specific star formation rates (sSFR). This was also shown in simulations done in \cite{Hirsch2017} and \cite{Hirsch2022}.

\section{ISM conditions at $z>5$} \label{ISM}

We compare our measurements of $\OIIIHb$ and $\NeOII$ to theoretical models to infer the physical conditions of the ISM. The models we compare to come from \citet{Kewl19} which uses both the  Starburst99 \citep{Leit99} models of stellar ionizing spectra and the MAPPINGS~V ionization code \citep{Suth18}. The Starburst99 model spectra includes mass loss and uses a \citet{salp55} IMF. The atomic data used in MAPPINGS~V comes from the CHIANTI~8 database \citep{dere97,delz15} which includes the effects of excitation, dust depletion, recombination and photoionization in the model $\HII$ regions. The ``Pressure Models'' of \citet{Kewl19} describe synthetic emission-line spectra created from different combinations of pressure $\log(P/k)$, ionization $\log(q)$, and metallicity $Z/Z_\odot$. These models interpolate between Starburst99 models and CHIANTI~8 data in order to match the grid. We use the following values of ionization and metallicity:
\begin{itemize}
    \item Ionization $\log(q) = [7,8,9]$, units of cm~s$^{-1}$
    \item Metallicity $Z/Z_\odot = [0.05,0.2,0.4,1.0]$
\end{itemize}
Due to neither emission line ratio varying significantly with pressure, we choose $\log{P}=8$ for the models. We also note the  $Z/Z_\odot = 0.05$ bin is extrapolated in the Starburst99 input spectra which causes those synthetic spectra to be the least certain of the theoretical predictions.

In Figure \ref{fig:Mapping} we compare our high redshift sample to the MAPPINGS~V models. The gray and blue histograms in both the \OIIIHb\ and \NeOII\ panel represent the NIRSpec and NIRCam WFSS emission-line ratio distribution, respectively. The four SMACS galaxies that have metallicity measurements are also included as red squares. These are compared to the MAPPINGS~V models that are represented by the three colored lines, where each color is a different ionization and each point is a different metallicity. The rest of the \OIIIHb\ ratios cover a range of moderate to high ionization. The \NeOII\ ratios are similarly well-matched to MAPPINGS~V models for moderate to high ionization.

Due to the degeneracy between ionization and metallicity in the MAPPINGS~V model prediction, it is difficult to infer ISM metallicity from the \OIIIHb\ ratio in Figure \ref{fig:Mapping}. To combat this we also make use of the OHNO diagram in Figure \ref{fig:O3Hb_ISM}, to give us a more narrow range of ISM conditions and highlight the evolution of ISM conditions with redshift. The gray contours represent the SDSS sample, the galaxies which are in the AGN region of the diagram do have higher ionization. The black points are the CLEAR sample from CLEAR. We include the measurements from the SMACS galaxies from \cite{Trump22} and the two NIRSpec AGN from \cite{Koce2022} as red squares and purple stars respectively. The $z=8.7$ AGN observed by \cite{Lars2023} is shown as a yellow star. Our sample is shown as pink circles representing the NIRSpec sample above $z>5$ and a pink bar for NIRCam WFSS. This bar represents the range of \OIIIHb\ covered by the WFSS sample as no \NeOII\ measurements were detected for these galaxies. The MAPPINGS~V models are represented the same way as in Figure \ref{fig:Mapping}. The inset arrow shows the typical scale and direction of $\sim$1~dex changes in metallicity. From this we can see our $z>5$ sample prefers a moderate to high ionization, $\log(q)=8,9$, with a moderate metallicity, $Z/Z_\odot=0.2,0.4$. The MAPPINGS~V models indicate that ionization increases from the $z \sim 0$ SDSS sample to the $z \sim 2$ CLEAR sample to the $z>5$ CEERS sample. We note that the three AGN sources have a higher \NeOII\ ratio and are best described by MAPPINGS~V models with higher ionization.

\section{Summary} \label{Summary}

In this work, we studied optical emission line ratios from $z\sim0$ to $z\sim9$ using SDSS, CLEAR, and CEERS data sets. We have used NIRCam WFSS to define two samples measuring $\Ha$ of 18 galaxies at $3.9<z<4.9$ and $\OIIIHb$ of 19 galaxies at $5.5<z<6.7$. This sample was found by first setting a constraint on the photometric redshifts, before visually inspecting for the emission lines. We also have three samples using NIRSpec observations to give additional measurements of $\Ha$ in 93 galaxies, $\OIIIHb$ of 96 galaxies, and a $\NeOII$ for 59 gal. The NIRSpec sample is selected with a $\mathrm{S/N} > 2$ in the emission lines of interest, before being visually inspected.

We studied these emission lines and summarize our results as follows:
\begin{itemize}

\item Our CEERS samples cover a redshift range of $2<z<9$ and show a significant ($3\sigma$) correlation between $\OIIIHb$ with redshift, $0.06\pm0.02$. There was also a marginal $2\sigma$ correlation with $\NeOII$ with redshift, $0.05\pm0.02$. When looking at the $\Ha$ SFR a very strong and significant correlation with a slope of $0.24\pm0.03$, with redshift was found.

\item We see a 0.33 dex increase in $\OIIIHb$, a 0.37 dex increase in $\NeOII$, and a 0.5 dex increase with \Ha\ SFR when comparing our CEERS $z>5$ sample to our $z\sim2$ sample from CLEAR .

\item We found \OIIIHb\ and \NeOII\ both have a significant correlations with \Hb\ SFR, with slopes of $0.26\pm0.03$ and $0.23\pm0.04$ respectively. \OIIIHb\ is also shown to have a significant anticorrelation with stellar mass with a slope of $-0.36\pm0.05$, while \NeOII\ has a marginal anticorrelation with a slope of $-0.2\pm0.07$.

\item The IllustrisTNG simulations match the measurements of our highest \OIIIHb\ values when we compare them to our observed galaxies at similar redshift, SFR, and stellar masses. 

\item When comparing our high redshift $z>5$ sample to MAPPINGS~V model spectra we found they are best described by high ionization with moderate metallicity. Comparing the CEERS line ratios with $z \sim 0$ and $z \sim 2$ samples indicates that the ISM ionization increases with increasing redshift.

\end{itemize}

These results indicate how emission-line ratios are evolving with redshift. We found that the galaxies' physical conditions at $z>5$ have higher SFRs and lower stellar mass. These galaxy properties are explained by higher ionization and lower metallicities. 

Larger samples of $z>5$ galaxies with JWST spectroscopy are needed to better disentangle the relationship between ISM conditions and galaxy properties at cosmic dawn. The NIRCam WFSS observations are very useful as a blind survey that includes galaxies missed by targeted NIRSpec observations, but the single-filter NIRCam WFSS observations in CEERS results in a limited wavelength range that includes only a single emission-line ratio for these galaxies. Multi-filter NIRCam WFSS observations would provide broader wavelength coverage and more effectively probe the ISM conditions for blind and representative samples of emission-line galaxies in the early Universe.

\section{Acknowledgements}

We acknowledge the work of our colleagues in the CEERS collaboration and everyone involved in the JWST mission. BEB and JRT acknowledge support from NASA grants JWST-ERS-01345, JWST-AR-01721, and NSF grant CAREER-1945546.

\bibliography{lib}{}

\end{document}